\documentclass[%
 preprint,superscriptaddress,
showpacs,preprintnumbers,
 amsmath,amssymb,
 aps,
 prc,
]{revtex4-1}
\usepackage{graphicx}
\usepackage{dcolumn}
\usepackage{bm}

\usepackage{subfigure}
\usepackage{multirow} 
\usepackage{mathrsfs} 
\begin{document}


\title{{\it Ab initio} nuclear many-body perturbation calculations in the Hartree-Fock basis}

\author{B. S. Hu (胡柏山)}
\affiliation{State Key Laboratory of
Nuclear Physics and Technology, School of Physics, Peking University, Beijing 100871,
China}
\author{F. R. Xu (许甫荣)} \thanks{frxu@pku.edu.cn}
\affiliation{State Key Laboratory of
Nuclear Physics and Technology, School of Physics, Peking University, Beijing 100871,
China}
\author{Z. H. Sun (孙中浩)}
\affiliation{State Key Laboratory of
Nuclear Physics and Technology, School of Physics, Peking University, Beijing 100871,
China}
\author{J. P. Vary}
\affiliation{Department of Physics and Astronomy, Iowa State University, Ames, Iowa 50011, USA}
\author{T. Li (李通)}
\affiliation{State Key Laboratory of
Nuclear Physics and Technology, School of Physics, Peking University, Beijing 100871,
China}


\date{\today}

\begin{abstract}
Starting from realistic nuclear forces,
the chiral N$^3$LO and JISP16,
we have applied many-body perturbation theory (MBPT)
to the structure of closed-shell nuclei,
$^4$He and $^{16}$O.
The two-body N$^3$LO interaction is softened by a similarity renormalization group transformation while JISP16 is adopted without renormalization.
The MBPT calculations are performed within the Hartree-Fock (HF) bases.
The angular momentum coupled scheme is used,
which can reduce the computational task.
Corrections up to the third order in energy
and up to the second order in radius are evaluated.
Higher-order corrections in the HF basis are small relative to the leading-order perturbative result.
Using the anti-symmetrized Goldstone diagram expansions of the wave function,
we directly correct the one-body density for the calculation of the radius,
rather than calculate corrections to
the occupation propabilities of single-particle orbits as found in other treatments.
We compare our results with other methods
where available and find good agreement.
This supports the conclusion that
our methods produce reasonably converged results with these interactions.
We also compare our results with experimental data.
\end{abstract}

\maketitle
\section{\label{sec:level1} Introduction}

A fundamental and challenging problem in nuclear structure theory
is the calculation of finite nuclei
starting from realistic nucleon-nucleon ($NN$) interactions.
The realistic nuclear forces, such as
CD-Bonn \cite{PhysRevC.63.024001},
Nijmegen \cite{PhysRevC.49.2950},
Argonne V18 (AV18) \cite{PhysRevC.51.38},
INOY \cite{PhysRevC.69.054001}
and chiral potential \cite{PhysRevC.68.041001,Machleidt20111},
contain strong short-range correlations
which cause convergence problems in the calculations of nuclear structures.
To deal with the strong short-range correlations and speed up the convergence,
realistic forces are usually processed by certain renormalizations.
A traditional approach is the G-matrix renormalization in the Brueckner-Bethe-Goldstone theory
\cite{PhysRev.97.1353,goldstone1957,PhysRev.129.225}
in which all particle ladder diagrams are summed.
Recently, a new class of renormalization methods has been developed,
including $V_{\text{low-}k}$ \cite{PhysRevC.65.051301,Bogner20031},
Similarity Renormalization Group (SRG) \cite{PhysRevC.75.061001},
Okubo-Lee-Suzuki \cite{okubo01111954,Suzuki01121980,Suzuki01071982,Suzuki01081983,Suzuki01121982,Suzuki01121994}
and Unitary Correlation Operator Method (UCOM) \cite{PhysRevC.72.034002,Roth201050}.
The renormalizations soften realistic $NN$ interactions
and generate effective Hamiltonians,
while all symmetries and observables are preserved in the low-energy domain.
The renormalization process also generates
effective multi-nucleon interactions (sometimes called "induced" interactions)
that are typically dropped for four or more nucleons interacting simultaneously.
 We will neglect three-nucleon and higher multi-nucleon interactions both "bare" and "induced".
There is another class of ``bare'' $NN$ forces which are
sufficiently soft that they can be used without renormalization,
e.g., the JISP interaction which is obtained by the $J$-matrix inverse scattering technique \cite{PhysRevC.70.044005,Shirokov200596,Shirokov200733}.
These interactions can often be used directly for nuclear structure calculations.

A renormalized $NN$ interaction should
retain its description of the experimental phase shifts
up to a cutoff.
At the same time,
the renormalized interaction provides better convergence
in nuclear structure calculations
without involving parameter refitting or additional parameters.
The calculations based on realistic forces are called {\it ab initio} methods
when they retain predictive power and accurate treatment of
the first principles of quantum mechanics.
There have been several {\it ab initio} many-body methods,
such as No-Core Shell Model (NCSM) \cite{PhysRevC.62.054311,PhysRevLett.84.5728,0954-3899-36-8-083101,Caprio2013179,Barrett2013131},
Green's Function Monte Carlo (GFMC) \cite{PhysRevC64014001,PhysRevC70054325,PhysRevC76064319,PhysRevC78065501}
and Coupled Cluster (CC) \cite{PhysRevLett101092502,PhysRevLett103062503,PhysRevC82034330}.
However, due to the limit of computer capability,
the NCSM and GFMC calculations are currently limited to light nuclei (e.g., $\leq ^{16}$O),
while the CC calculations are limited to nuclei near double closed shells.

While renormalization methods typically address
short-range correlations,
the Hartree-Fock (HF) approach is used to treat long-range correlations.
However, the conventional HF method that takes only one Slater determinant
describes the motion of nucleons in the average field of other nucleons and
neglects higher-order correlations.
For a phenomenological potential,
one can adjust parameters to
improve the agreement of the HF results with data.
For realistic $NN$ interactions,
one needs to go beyond the HF approach
to include the intermediate-range correlations
which are missing in the lowest order HF approach.
The many-body perturbation theory (MBPT) is a powerful tool
to include the missing correlations \cite{bartlett2009,PhysRevC.68.034320,PhysRevC.69.034332,PhysRevC.73.044312}.
The perturbation method starts from a solvable mean-field problem
and derives a correlated perturbed solution.
The most well-known perturbation expansions are the Brillouin-Wigner (BW) \cite{Brillouin1932,Wigner1935} and Rayleigh-Schr{\"o}dinger (RS) \cite{Rayleigh1894,Schrodinger1926} methods.
MBPT calculations are usually performed with an order-by-order expansion represented in the form of groups of diagrams \cite{bartlett2009}.
The diagrams of MBPT proliferate as one goes to higher orders but some techniques, such as those introduced by Bruekner \cite{PhysRev.100.36}, lead to useful cancellations of entire classes of diagrams.
This leads to the linked-diagram theorem which simplifies greatly perturbation calculations up to high orders.
Goldstone first proved the theorem valid to all orders in the non-degenerate case \cite{goldstone1957}.
Later, the theorem was extended to the degenerate case \cite{RevModPhys39771,Johnson1971172,Kuo197165,sandar1969}.
The linked-diagram theorem in the degenerate case is often referred to as the folded-diagram method.

Some recent works \cite{PhysRevC.68.034320,PhysRevC.69.034332,PhysRevC.73.044312} show that
the MBPT corrections to HF can significantly improve calculations
which were based on realistic forces.
The authors used different renormalization schemes,
$V_{\text{low-}k}$, OLS and UCOM,
and obtained the convergence of low-order MBPT calculations \cite{PhysRevC.68.034320,PhysRevC.69.034332,PhysRevC.73.044312}.
In the present work,
we perform similar MBPT calculations with the SRG-renormalized chiral N$^3$LO potential \cite{PhysRevC.68.041001,Machleidt20111}
and the ``bare'' JISP16 interaction \cite{PhysRevC.70.044005,Shirokov200596,Shirokov200733}.
We also calculate the MBPT corrections to the nuclear radius
with the anti-symmetrized Goldstone (ASG) diagrams of
the one-body density (up to the second order).
We note that, in Ref.~\cite{PhysRevC.68.034320},
the same ASG diagrams for the corrections to energy
were used for the corrections to the radius.
In Refs. \cite{PhysRevC.69.034332,PhysRevC.73.044312},
corrections to the radius were approximated through corrections to occupation probabilities.
In order to reduce computational task,
we calculate the diagrams in the angular momentum coupling representation.
Our MBPT corrections to energy are up to the third order,
while our MBPT corrections to the radius are up to the second order.

\section{\label{sec:1}Theoretical framework}

\subsection{ The effective Hamiltonian }
The intrinsic Hamiltonian of the $A$-nucleon system used in this work reads
\begin{equation}
\begin{array}{ll}
\text{\^{H}}= \displaystyle\sum_{i<j}^{A} \frac{(\vec{p}_{i}-\vec{p}_{j})^{2}}{2mA} +
\displaystyle\sum_{i<j}^{A}   V_{NN,ij},
\end{array}
\label{eq1}
\end{equation}
where the notation is standard.
The first term on the right is the intrinsic kinetic energy,
and $V_{NN,ij}$ is the $NN$ interaction
including the Coulomb interaction between the protons.
We do not include a three-body interacton.
In the present work,
two different $NN$ interactions have been adopted for comparison.
One is the chiral potential N$^3$LO developed by Entem and Machleidt \cite{PhysRevC.68.041001}.
Another one is the ``bare'' interaction JISP16 \cite{PhysRevC.70.044005,Shirokov200596,Shirokov200733}.

The N$^3$LO potential is renormalized by using the SRG technique
to soften the short-range repulsion and short-range tensor components.
The SRG method is based on a continuous unitary transformation
that suppresses off-diagonal matrix elements
and drives the Hamiltonian towards a band-diagonal form \cite{PhysRevC.75.061001}.
The process leads to high- and low-momentum parts of the Hamiltonian being decoupled.
This implies that the renormalized potential becomes softer
and more perturbative than the original one.
In principle, the SRG method generates three-body, four-body, etc., effective interactions.
We neglect these induced terms for the purposes of examining the similarities and differences of results with NN interactions alone.
After the renormalization, the Coulomb interaction between protons is added.

The ``bare'' JISP16 interaction is obtained
by the phase-equivalent transformations of the $J$-matrix inverse scattering potential.
The parameters are determined by
fitting to not only the $NN$ scattering data
but also the binding energies and spectra of nuclei with $A\leq16$ \cite{Shirokov200733}.
In the JISP16 potential,
the off-shell freedom is exploited
to improve the description of light nuclei
by phase-equivalent transformations.
Polyzou and Glockle \cite{latePolyzou} have shown that
changing the off-shell properties of the two-body potential
is equivalent to adding many-body interactions.
Therefore, the phase-equivalent transformation
can minimize the need of three-body interactions.
The ``bare'' JISP16 interaction has been used extensively and successfully in configuration interaction calculations of light nuclei \cite{Maris:2013poa,Shirokov:2014}
and in nuclear matter \cite{Shirokov:2014kqa}.

\subsection{Spherical Hartree-Fock formulation}

With the effective Hamiltonian established,
we first perform the HF calculation and then calculate the MBPT corrections to the HF result.
For simplicity of computational effort, we limit our investigations here to the spherical, closed-shell, nuclei $^4$He and $^{16}$O.
These systems are sufficient to gain initial insights into the convergence rates of the ground-state energy and radius with these realistic interactions.

The spherical symmetry preserves
the quantum numbers of the orbital angular momentum ($l$), the total angular momentum ($j$)
and its projection ($m_j$) for the HF single-particle states.
In the spherical harmonic oscillator (HO) basis $|n l j m_j m_t \rangle$,
the HF single-particle state $|\alpha \rangle$ can be written as
\begin{equation}
| \alpha \rangle=| \nu l j m_{j} m_{t} \rangle
=\displaystyle\sum_{n} D_{n}^{(\nu ljm_{j}m_{t})} | n l j m_{j} m_{t} \rangle,
\end{equation}
where the labels are standard with $n$ and $m_t$
for the radial quantum number of the HO basis and isospin projection, respectively.
The HF wave function for the $A$-body nucleus is then represented by
an anti-symmetrized Slater determinant constructed
with the HF single-particle states.
By varying the HF energy expectation value
(with respect to the coefficients $D_n^{(\nu l j m_j m_t)}$),
we obtain the HF single-particle eigen equations,
\begin{eqnarray}
\displaystyle\sum_{n_{2}}
h_{n_{1}n_{2}}^{(ljm_{j}m_{t})}D_{n_{2}}^{(\nu ljm_{j}m_{t})}
=\varepsilon_{\nu ljm_{j}m_{t}}D^{(\nu ljm_{j}m_{t})}_{n_{1}},
\end{eqnarray}
where $\varepsilon_{\nu l j m_j m_t}$ represents
the HF single-particle eigen energies,
and $h_{n_{1} n_{2}}^{(l j m_{j} m_{t})}$ designates the matrix elements of the HF single-particle Hamiltonian given by
\begin{eqnarray}
h_{n_{1}n_{2}}^{(ljm_{j}m_{t})}=
\displaystyle\sum_{l'j'm'_{j}m'_{t}}
\displaystyle\sum_{n'_{1}n'_{2}}
H_{n_{1}n'_{1}n_{2}n'_{2}}^{(ljm_{j}m_{t};l'j'm'_{j}m'_{t})}
\rho_{n'_{1}n'_{2}}^{(l'j'm'_{j}m'_{t})},
\label{hfsph}
\end{eqnarray}
where $H^{(l j m_j m_t, l^\prime j^\prime m_j^\prime m_t^\prime)}_{n_1 n_1^\prime n_2 n_2^\prime}$ and $\rho^{( l^\prime j^\prime m_j^\prime m_t^\prime)}_{ n_1^\prime n_2^\prime}$ are the matrix elements of the two-body effective Hamiltonian $\hat{H}$ and one-body density, respectively.
They can be written
\begin{eqnarray}
H_{n_{1}n'_{1}n_{2}n'_{2}}^{(ljm_{j}m_{t};l'j'm'_{j}m'_{t})}=
\langle n_{1}ljm_{j}m_{t},n'_{1}l'j'm'_{j}m'_{t}|\hat{H} |
n_{2}ljm_{j}m_{t},n'_{2}l'j'm'_{j}m'_{t}\rangle
\end{eqnarray}
and
\begin{eqnarray}
\rho_{n'_{1}n'_{2}}^{(l'j'm'_{j}m'_{t})}=
\displaystyle\sum_{u}\mathscr{N}^{(ul'j'm'_{j}m'_{t})}
D_{n'_{1}}^{\ast (ul'j'm'_{j}m'_{t})}D_{n'_{2}}^{(ul'j'm'_{j}m'_{t})},
\end{eqnarray}
where $\mathscr{N}^{(\mu l^\prime j^\prime m_j ^\prime m_t^\prime)}$
is the occupation number of the HF single-particle orbit,
i.e., $\mathscr{N}^{(\mu l^\prime j^\prime m_j ^\prime m_t^\prime)}
=1$ (occupied) or $0$ (unoccupied).

In practice,
we diagonalize the following equation to solve the HF single-particle eigenvalue problem
\begin{eqnarray}
\displaystyle\sum_{n_{2}} \left[ \displaystyle\sum_{n'_{1}n'_{2}}
\displaystyle\sum_{l'j'm'_{j}m'_{t}}
H^{ (ljm_{j}m_{t},l'j'm'_{j}m'_{t})}_{n_{1}n'_{1},n_{2}n'_{2}}
\rho^{ ( l'j'm'_{j}m'_{t})}_{n'_{1}n'_{2}} \right]
D^{(\nu ljm_{j}m_{t})}_{n_{2} }
=\varepsilon_{\nu ljm_{j}m_{t}}D^{(\nu ljm_{j}m_{t})}_{n_{1}}.
\label{hfe}
\end{eqnarray}
This is a nonlinear equation
with respect to variational coefficients $D_{n}^{(\nu l j m_{j} m_{t})}$.
In the spherical closed shell,
the HF single-particle eigenvalues are independent of the magnetic quantum number $m_j$,
which leads to a $2j+1$ degeneracy.
In this case, we can rewrite the eigenvalues by omitting $m_{j}$,
i.e., $ D_{n}^{(\nu l j m_{t})}=D_n^{(\nu l j m_{j} m_{t})}$
and $\varepsilon_{\nu l j  m_{t}}=\varepsilon_{\nu l j m_{j} m_{t}}$.
Then we can simplify  Eq.~(\ref{hfe})
in the angular momentum coupled representation as follows\cite{PhysRevC.73.044312},
\begin{equation}
\begin{array}{ll}
\displaystyle\sum_{n_{2}} \Bigg[ \displaystyle\sum_{n'_{1}n'_{2}}
\displaystyle\sum_{l'j'm'_{t}} \displaystyle\sum_{J}
\frac{2J+1}{(2j+1)(2j'+1)}
\sqrt{1+\delta_{k_{1}k'_{1}}}
\sqrt{1+\delta_{k_{2}k'_{2}}}
\\ \times
\left\langle n_{1}ljm_{t}, n'_{1}l'j'm'_{t};J|\text{\^{H}}|n_{2}ljm_{t}, n'_{2}l'j'm'_{t};J\right\rangle
\rho^{ (l'j'm'_{t})}_{n'_{1}n'_{2}} \Bigg]
\\ \times
D^{(\nu ljm_{t})}_{n_{2}}
=\varepsilon_{\nu ljm_{t}}D^{(\nu ljm_{t})}_{n_{1}}
\end{array}
\end{equation}
with $\delta_{kk'}=\delta_{nn'}\delta_{ll'}\delta_{jj'}\delta_{m_{t}m'_{t}}$ and
one-body density matrix
\begin{equation}
\begin{array}{ll}
\rho^{ (l'j'm'_{t})}_{n'_{1}n'_{2}}=
\displaystyle\sum_{\mu}O^{(\mu l'j'm'_{t})}
D^{ \text{\textasteriskcentered} (\mu l'j'm'_{t}) }_{n'_{1} } D^{ (\mu l'j'm'_{t})}_{n'_{2} },
\end{array}
\end{equation}
where $\displaystyle O^{(\mu l'j'm'_{t})}$ is the number of the occupied magnetic subshell,
i.e., $O^{(\mu l'j'm'_{t})}=2j'+1$ (occupied) or 0 (unoccupied).

\subsection{ Rayleigh-Schr\"{o}dinger perturbation theory}

We can separate the $A$-nucleon Hamiltonian Eq.~(\ref{eq1}) into a zero-order part $\hat{H_{0}}$ and a perturbation $\hat{V}$,
\begin{equation}
\text{\^{H}}=\text{\^{H}}_{0}+(\text{\^{H}}-\text{\^{H}}_{0})=\text{\^{H}}_{0}+\text{\^{V}}.
\end{equation}
The exact solutions of the $A$-nucleon system are
\begin{equation}
\text{\^{H}}\Psi_{n}=E_{n} \Psi_{n}, \qquad n=0,1,2,...
\end{equation}
For the zero-order part, we write
\begin{equation}
\text{\^{H}}_{0} \Phi_{n}=E^{(0)}_{n} \Phi_{n}, \qquad n=0,1,2,...
\end{equation}

If we choose the HF single-particle Hamiltonian Eq.~(\ref{hfsph})
as $H_{0}$,
the zero-order energy $E_{0}^{(0)}$ is simply the summation of the
single-particle energies up to the Fermi level.
In the present work,
we only investigate the ground states of closed-shell nuclei.
For simplicity, we denote the ground-state energy $E_{0}$
and wave function $\Psi_{0}$ by $E$ and $\Psi$, respectively,
omitting the subscript.
For the ground state $(n=0)$,
we formulate the Rayleigh-Schr{\"o}dinger perturbation theory (RSPT), as follows,
\begin{equation}
\chi= \Psi - \Phi_{0},
\end{equation}
\begin{equation}
\Delta E= E- E^{(0)},
\label{r0}
\end{equation}
\begin{equation}
\Psi= \displaystyle\sum_{m=0}^{\infty} \big{[} \text{\^{R}}_{0}(E^{(0)}) (\text{\^{V}}-\Delta E)\big{]}^{m} \Phi_{0},
\label{r1}
\end{equation}
\begin{equation}
\Delta E= \displaystyle\sum_{m=0}^{\infty} \langle \Phi_{0} |\text{\^{V}} \big{[}\text{\^{R}}_{0}(E^{(0)})
(\text{\^{V}}-\Delta E)\big{]}^{m} |\Phi_{0}\rangle,
\label{r2}
\end{equation}
where  $\text{\^{R}}_{0}=\displaystyle\sum_{i \neq 0} \frac{|\Phi_{i}\rangle\langle\Phi_{i}|}{E_{0}^{(0)}-E_{i}^{(0)} }$
is called the resolvent of $\text{\^{H}}_{0}$. Here we use intermediate normalization
\begin{equation}
\begin{array}{ll}
\langle\Phi_{n}|\Phi_{n}\rangle=1, \qquad \langle\chi_{n}|\Phi_{n}\rangle=0,
\\
\langle\Psi_{n}|\Phi_{n}\rangle=1, \qquad
\langle\Psi_{n}|\Psi_{n}\rangle=1+\langle\chi_{n}|\chi_{n}\rangle.
\end{array}
\label{norm}
\end{equation}
Arranging the above expressions according to the perturbation orders of $\text{\^{V}}$,
we have
\begin{equation}
E=E^{(0)}+E^{(1)}+E^{(2)}+E^{(3)}+\ldots
\end{equation}
The first-, second-, third-order corrections are
\begin{equation}
E^{(1)}=\langle \Phi_{0} |\text{\^{V}} |\Phi_{0}\rangle,
\end{equation}
\begin{equation}
E^{(2)}=\langle \Phi_{0} |\text{\^{V}} \text{\^{R}}_{0} \text{\^{V}} |\Phi_{0}\rangle,
\end{equation}
\begin{equation}
E^{(3)}=\langle \Phi_{0} |\text{\^{V}} \text{\^{R}}_{0}
(\text{\^{V}}-\langle \Phi_{0} |\text{\^{V}} |\Phi_{0}\rangle) \text{\^{R}}_{0} \text{\^{V}} |\Phi_{0}\rangle.
\end{equation}
Similarly, the wave function can be written in the perturbation scheme
\begin{equation}
\Psi=\Phi_{0}+\Psi^{(1)}+\Psi^{(2)}+\ldots
\end{equation}
with
\begin{equation}
\Psi^{(1)}=\text{\^{R}}_{0} \text{\^{V}} |\Phi_{0}\rangle
\end{equation}
and
\begin{equation}
\Psi^{(2)}=\text{\^{R}}_{0} (\text{\^{V}}-E^{(1)})\text{\^{R}}_{0} \text{\^{V}} |\Phi_{0}\rangle
\end{equation}
for the first- and second-order corrections to the
wave function, respectively.
We can use the diagrammatic approach to describe various terms in RSPT.
The ASG diagrams are the most commonly-used method of the diagrammatic representation.

\subsection{ Diagrammatic expansion for Rayleigh-Schr\"{o}dinger perturbation theory in the Hartree-Fock basis}

If we choose the HF Hamiltonian as an auxiliary zero-order one-body Hamiltonian $\text{\^{H}}_{0}$,
many of the ASG diagrams are cancelled \cite{bartlett2009}.
Only a small number of low-order ASG diagrams for RSPT remain.
In this subsection, we give the remaining AGS diagrams for the energy and wave function
written in the standard perturbation theory \cite{kuo1990}.
We consider corrections up to third order for the energy
and second order for the wave function.
To evaluate other observables that can be expressed by one-body operators,
we calculate the corrections up to second order for the one-body density.
It has been shown that
the corrections up to third order for the energy
in the HF basis give well-converged results for soft interactions \cite{Tichai:2016joa}.
Spherical HF (SHF) produces degenerate single-particle states,
so we can evaluate the vacuum-to-vacuum linked diagrams
in angular momentum coupled representation \cite{Kuo1981237}
which is computationally efficient.
\begin{figure}
\setlength{\fboxrule}{0.7pt}
\fbox{
\shortstack[c]{
\subfigure[]
{
\label{fig:subfig:1} 
\includegraphics[width=1.6in]{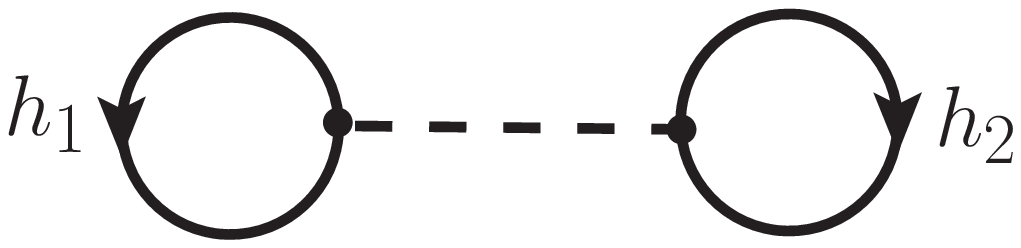}
}
\\
\subfigure[]
{
\label{fig:subfig:2} 
\includegraphics[width=1.5in]{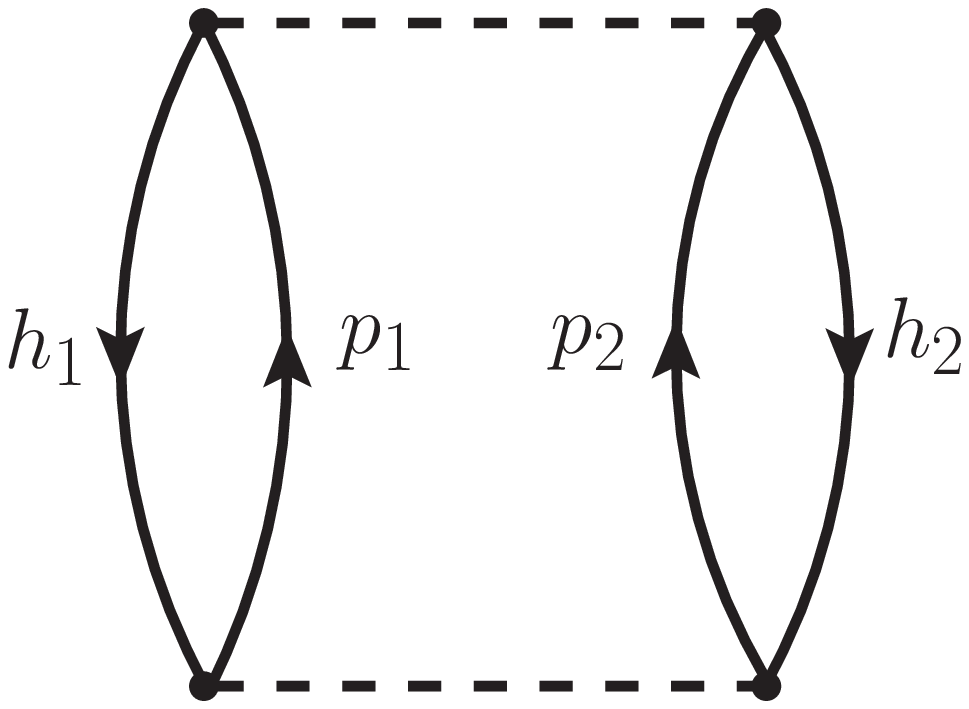}
}
\subfigure[]
{
\label{fig:subfig:3} 
\includegraphics[width=1.5in]{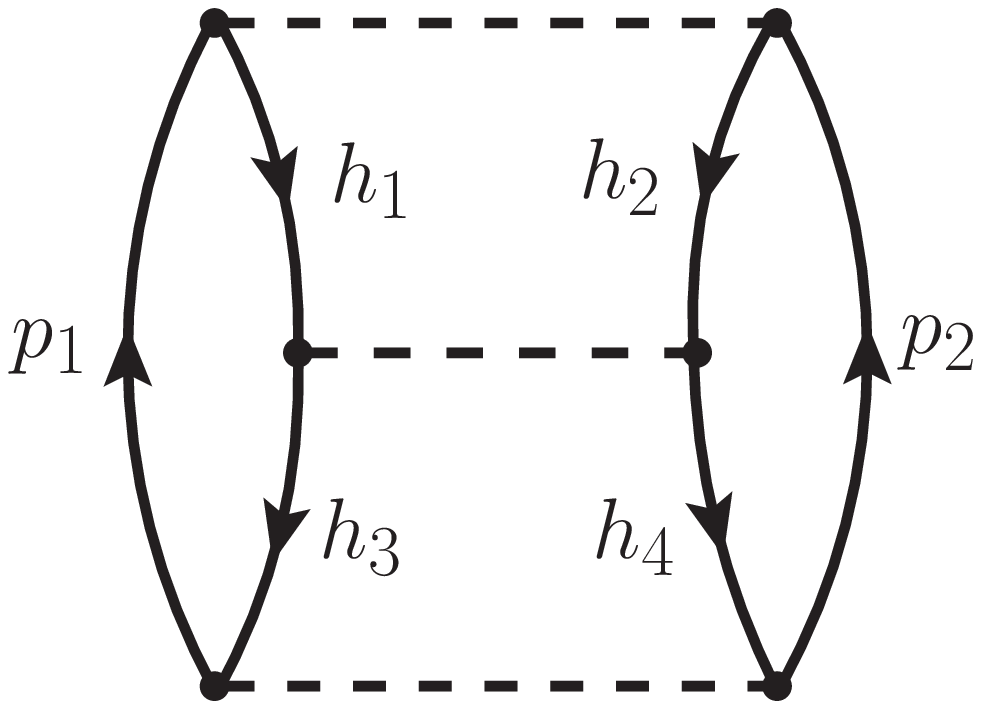}
}
\\
\subfigure[]
{
\label{fig:subfig:4} 
\includegraphics[width=1.5in]{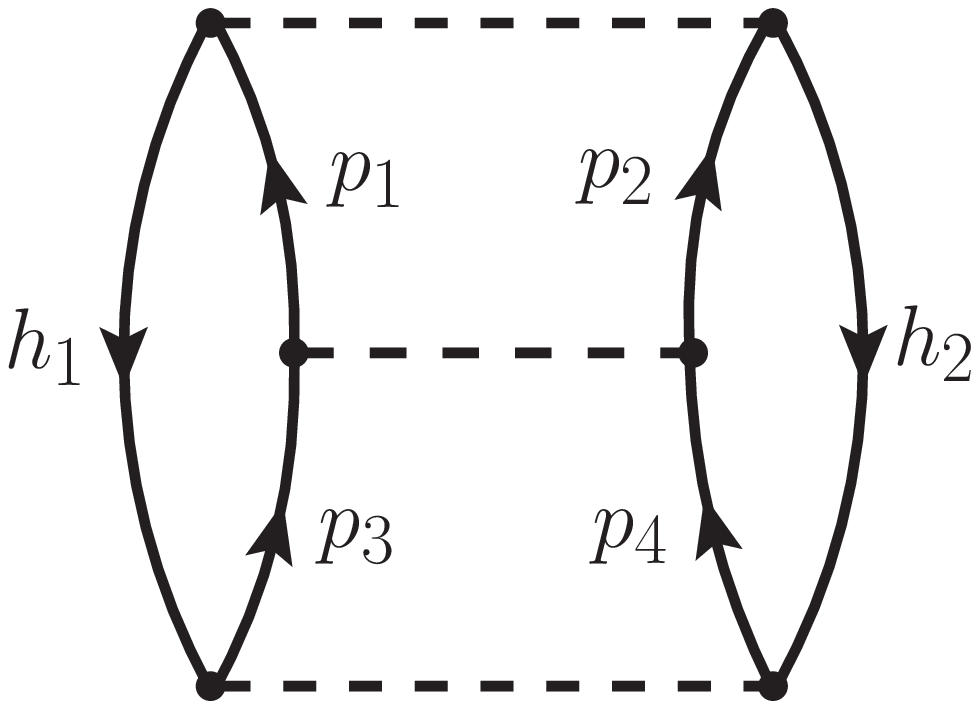}
}
\subfigure[]
{
\label{fig:subfig:5} 
\includegraphics[width=1.5in]{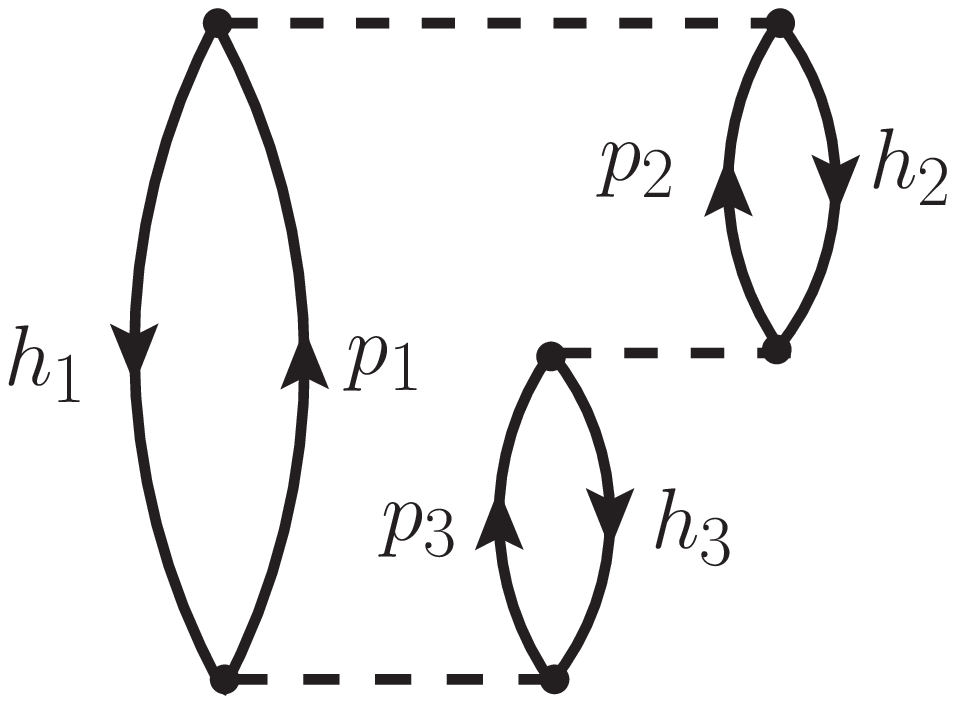}
}
}
}
\caption{The first-, second-, and third-order ASG diagrams
of energy corrections in the RS expansion \cite{PhysRevC.68.034320}.}
\label{fig:1} 
\end{figure}

Fig.~\ref{fig:1} displays the ASG diagrams corresponding to
the first-, second- and third-order corrections to the energy in RSPT.
The vertices, i.e., the dashed lines, represent $\hat H$ in Eq.~(\ref{eq1}). The diagrams (a) and (b) are for $E^{(1)}$ and $E^{(2)}$, respectively, while the diagrams (c), (d) and (e) sum up for $E^{(3)}$.
The zero-order energy $E^{(0)}$ is the simple summation of the HF single-particle energies up to the Fermi level,
i.e., $E^{(0)}=\displaystyle\sum_{i=1}^{A} \varepsilon_{i}$,
where $\varepsilon_{i}$ represents the HF single-particle energy.
The summation of the $E^{(0)}$ and $E^{(1)}$ gives the HF energy,
i.e., $E_{\text{HF}}=E^{(0)}+E^{(1)}=\dfrac{1}{2}\displaystyle\sum_{i=1}^{A} \varepsilon_{i}$,
since the initial Hamiltonian is entirely expressed in relative coordinates
\cite{PhysRevC.37.1240,PhysRevC.69.034332}.

\subsubsection{Corrections to the one-body density}

MBPT corrections to the wave function bring configuration mixing.
The convergence can be discussed in order-by-order perturbation calculations.
Any observable that is expressed by one-body operators
can be calculated by using the One-Body Density Matrix (OBDM).
By definition, the local one-body density operator
in an $A$-body Hilbert space is written as \cite{PhysRevC.86.034325}
\begin{equation}
\hat{\rho} (\vec{r})=\displaystyle\sum_{k=1}^{A}\delta^{3} \left(\vec{r}- \vec{r}_{k} \right)
=\displaystyle\sum_{k=1}^{A} \frac{\delta \left(r- r_{k} \right) }{r^{2}}
\displaystyle\sum_{lm} Y_{lm}^{\ast}(\text{\^{r}}_{k}) Y_{lm}(\text{\^{r}}),
\end{equation}
where $\text{\^{r}}$ is the unit vector in the direction $\vec{r}$,
and $Y_{lm}(\text{\^{r}})$ is the spherical harmonic function.

We can write the density operator in the second quantization representation
in the HO basis as
\begin{eqnarray}
\hat{\rho}(\vec{r})
=&&\displaystyle\sum_{K}
\displaystyle\sum_{n_{1}l_{1}j_{1}}
\displaystyle\sum_{n_{2}l_{2}j_{2}}\displaystyle\sum_{m_{j}}
R_{n_{1}l_{1}}(r)R_{n_{2}l_{2}}(r)
\frac{-Y_{K0}^{\ast}(\text{\^{r}})}{\sqrt{2K+1}}
\nonumber \\ &&\times
\left\langle l_{1} \frac{1}{2} j_{1} \left|| Y_{K} | \right| l_{2} \frac{1}{2} j_{2}  \right\rangle
\left\langle j_{1} m_{j} j_{2} -m_{j} | K 0  \right\rangle
\nonumber \\ &&\times
 (-1)^{j_{2}+m_{j}}
a_{n_{1}l_{1}j_{1}m_{j}}^{\dag}a_{n_{2}l_{2}j_{2}m_{j}}
\end{eqnarray}
with
\begin{eqnarray}
\left\langle l_{1} \frac{1}{2} j_{1} \left|| Y_{K} | \right| l_{2} \frac{1}{2} j_{2}  \right\rangle
=&& \frac{1}{\sqrt{4 \pi}} \text{\^{j}}_{1}\text{\^{j}}_{2}\text{\^{l}}_{1}\text{\^{l}}_{2}(-1)^{j_{1}+\frac{1}{2}}
\left\langle l_{1} 0 l_{2} 0 | K 0  \right\rangle
\nonumber \\ &&\times
\left\{
  \begin{array}{ccc}
    j_{1} & j_{2} &  K\\
    l_{2} & l_{1} &  \frac{1}{2}\\
  \end{array}
\right\}.
\end{eqnarray}
The $R_{nl}$'s are the radial components of the HO wave function.
We use the Condon-Shortley convention for the Clebsch-Gordan coefficients.
Since we are dealing with a spherically symmetric system (K=0),
we can obtain a simple form,
\begin{equation}
\hat{\rho} (\vec{r})=\displaystyle\sum_{n_{1}n_{2}}\displaystyle\sum_{ljm_{j}}
\left[ \frac{R_{n_{1}l}(r)R_{n_{2}l}(r)}{4 \pi} \right]
a_{n_{1}ljm_{j}}^{\dag}a_{n_{2}ljm_{j}}.
\end{equation}

By introducing the normally-ordered product relative to the SHF ground state $|\Phi_{0} \rangle$,
the local one-body density operator can be written as
\begin{equation}
\hat{\rho} (\vec{r})= \rho_{0}(\vec{r}) +\hat{\rho}_{N}
=\rho_{0}(\vec{r}) +
\displaystyle\sum_{i,j} \rho_{ij}:c_{i}^{\dag}c_{j}:,
\label{normal-order}
\end{equation}
where $\rho_{0}(\vec{r})=\langle \Phi_{0}| \hat{\rho} (\vec{r}) | \Phi_{0}\rangle$
gives the HF density,
while $\hat{\rho}_{N}=\displaystyle\sum_{i,j} \rho_{ij}:c_{i}^{\dag}c_{j}:$
brings corrections to the density.
$\rho_{ij}$ is the density matrix elements $\langle i|\rho (\vec{r})|j \rangle$,
and $:c_{i}^{\dag}c_{j}:$ indicates the normally-ordered product of
the creation and annihilation operators.
It is required that all annihilation and creation operators
which take $|\Phi_{0}\rangle$ to zero when acting on it are to the
right of all other operators which do not take $|\Phi_{0}\rangle$ to zero.
The expectation value of the density is obtained
with the corrected wave function through Eq.~(\ref{normal-order}).
In the present work,
we consider the first- and second-order wave function corrections.
\begin{widetext}
\begin{figure}
\setlength{\fboxrule}{0.6pt}
\setlength{\fboxsep}{0.15cm}
\fbox{
\shortstack[c]{
\subfigure[]
{
\includegraphics[width=1.36in]{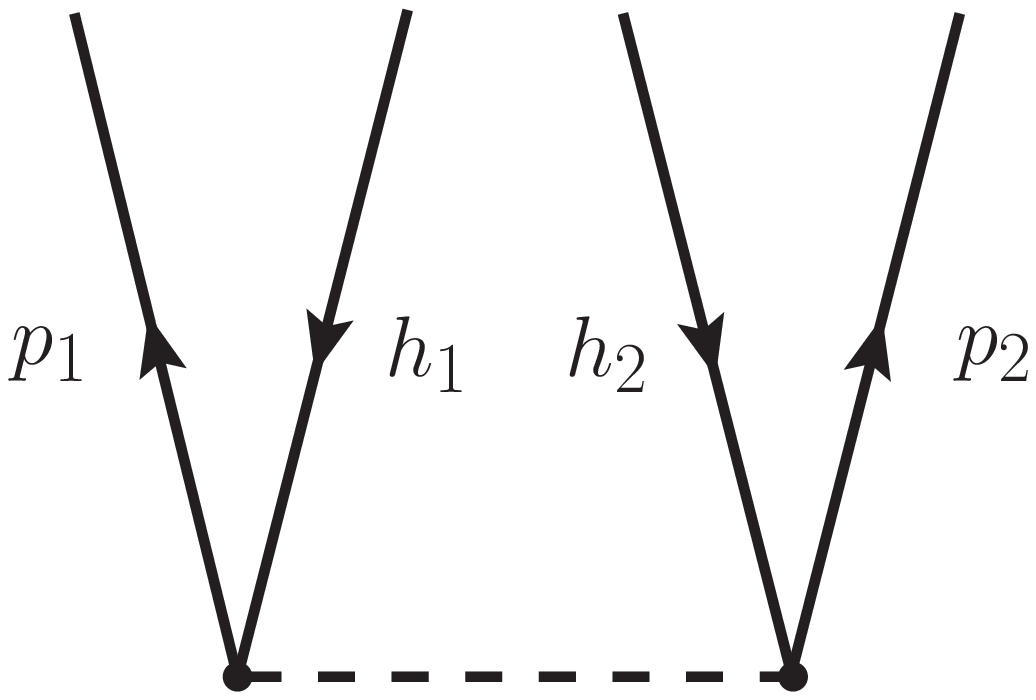}
}
\subfigure[]
{
\includegraphics[width=1.34in]{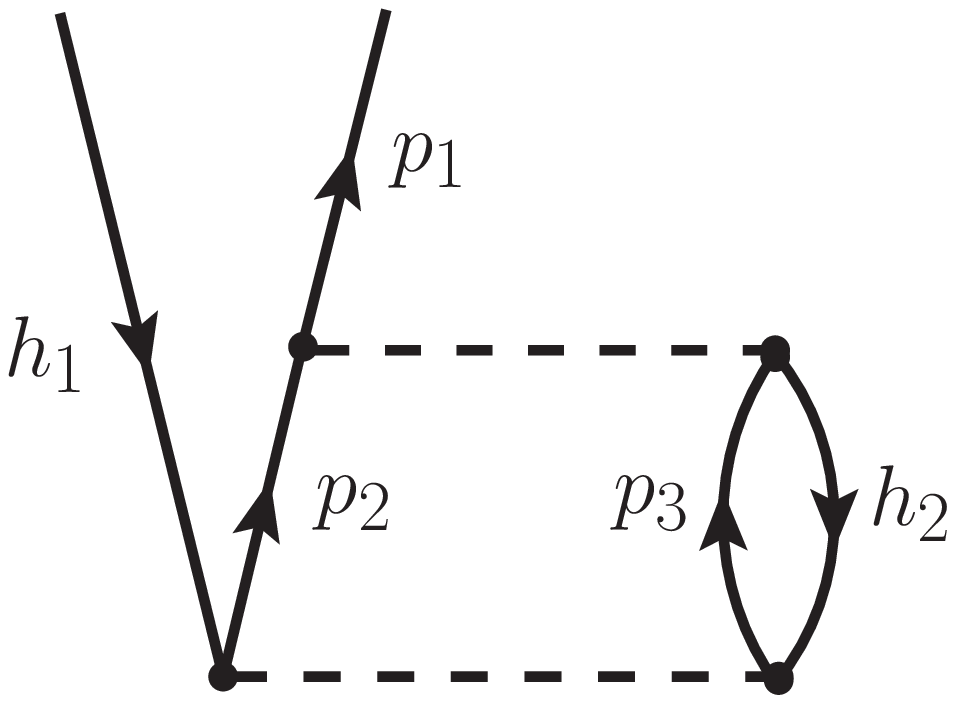}
}
\subfigure[]
{
\includegraphics[width=1.34in]{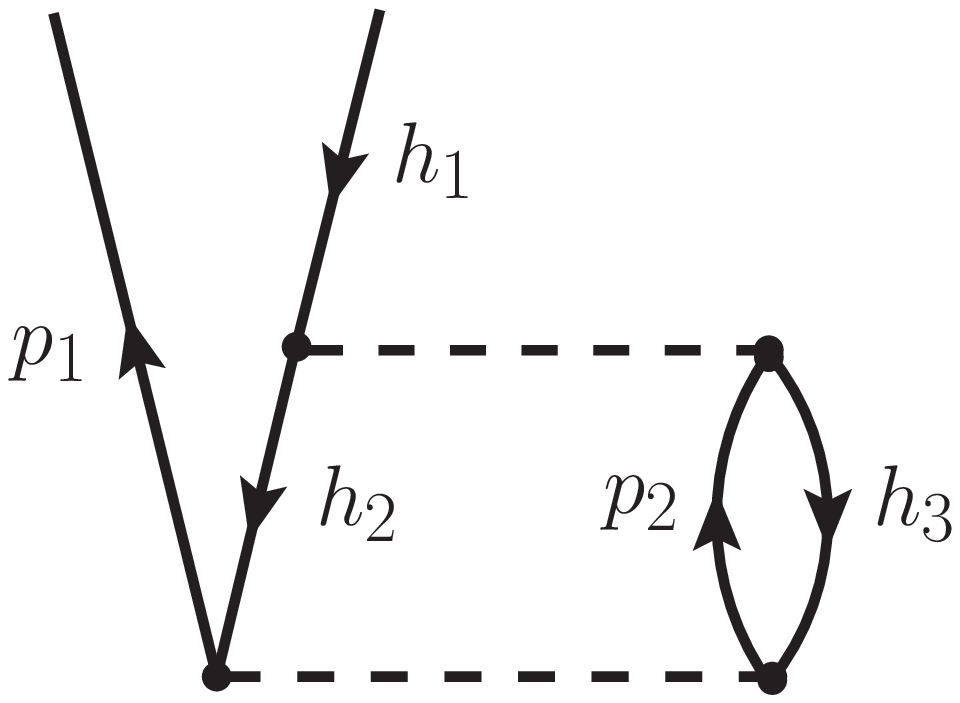}
}
\\
\subfigure[]
{
\includegraphics[width=1.34in]{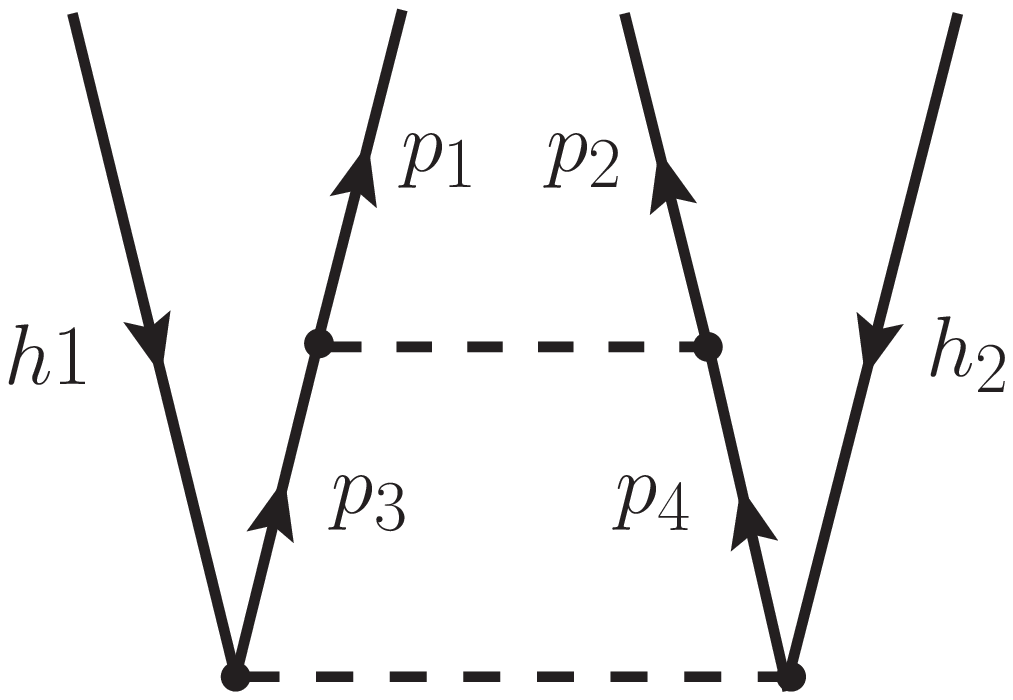}
}
\subfigure[]
{
\includegraphics[width=1.34in]{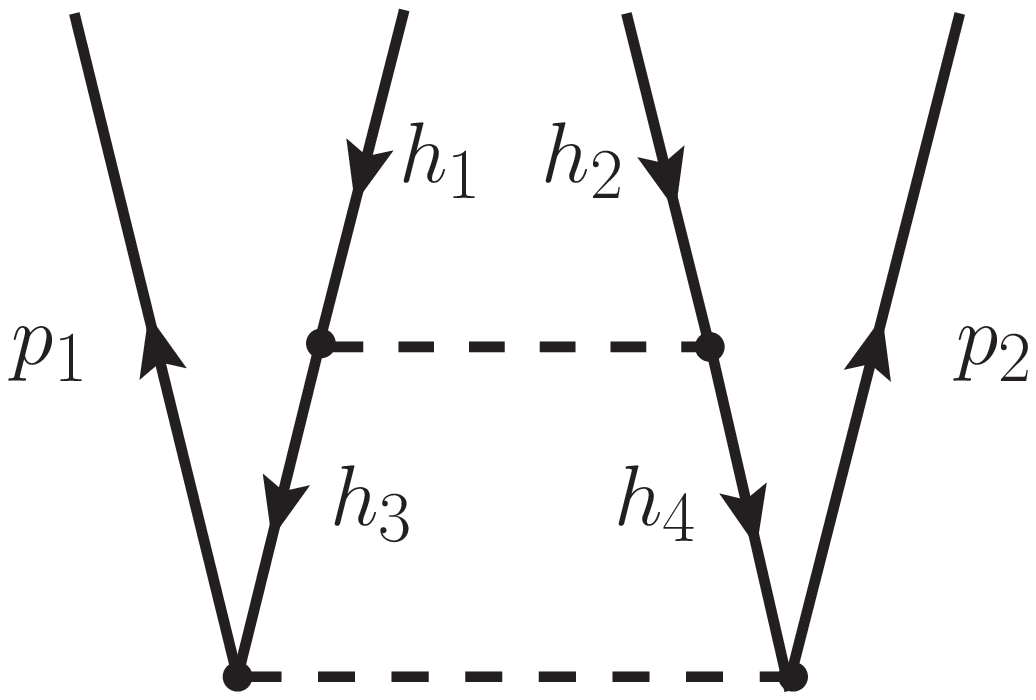}
}
\subfigure[]
{
\includegraphics[width=1.35in]{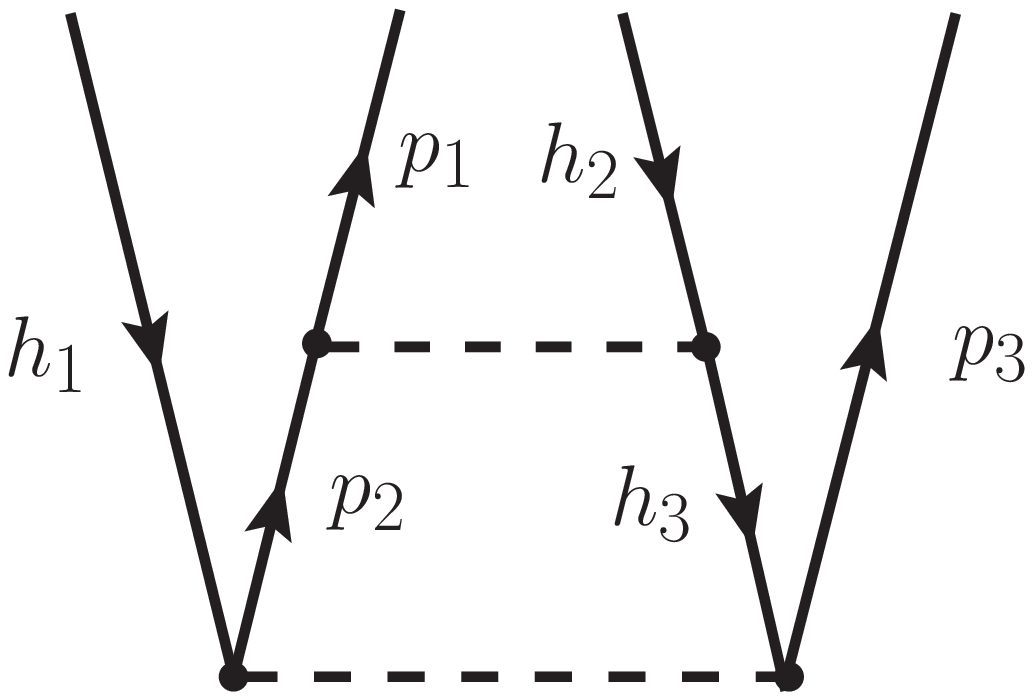}
}
\\
\subfigure[]
{
\includegraphics[width=1.81in]{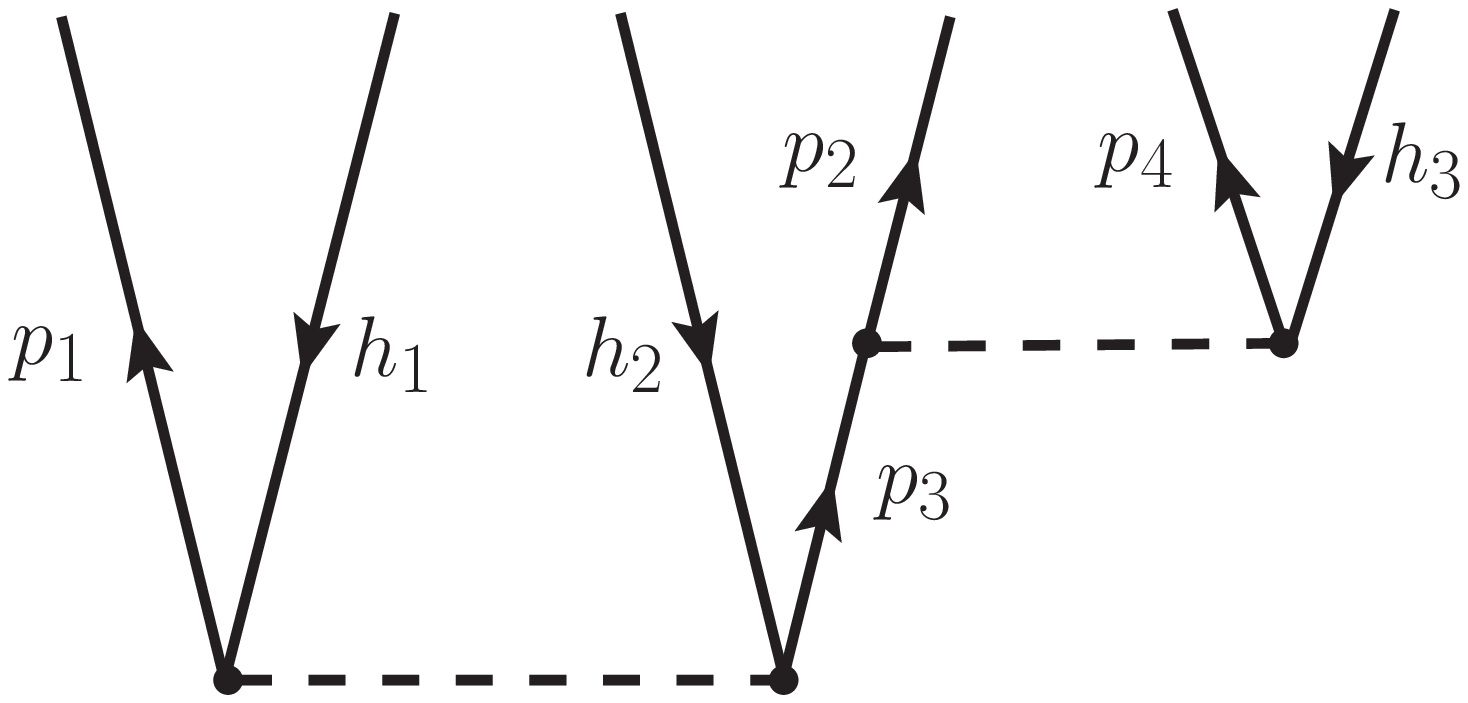}
}
\subfigure[]
{
\includegraphics[width=1.81in]{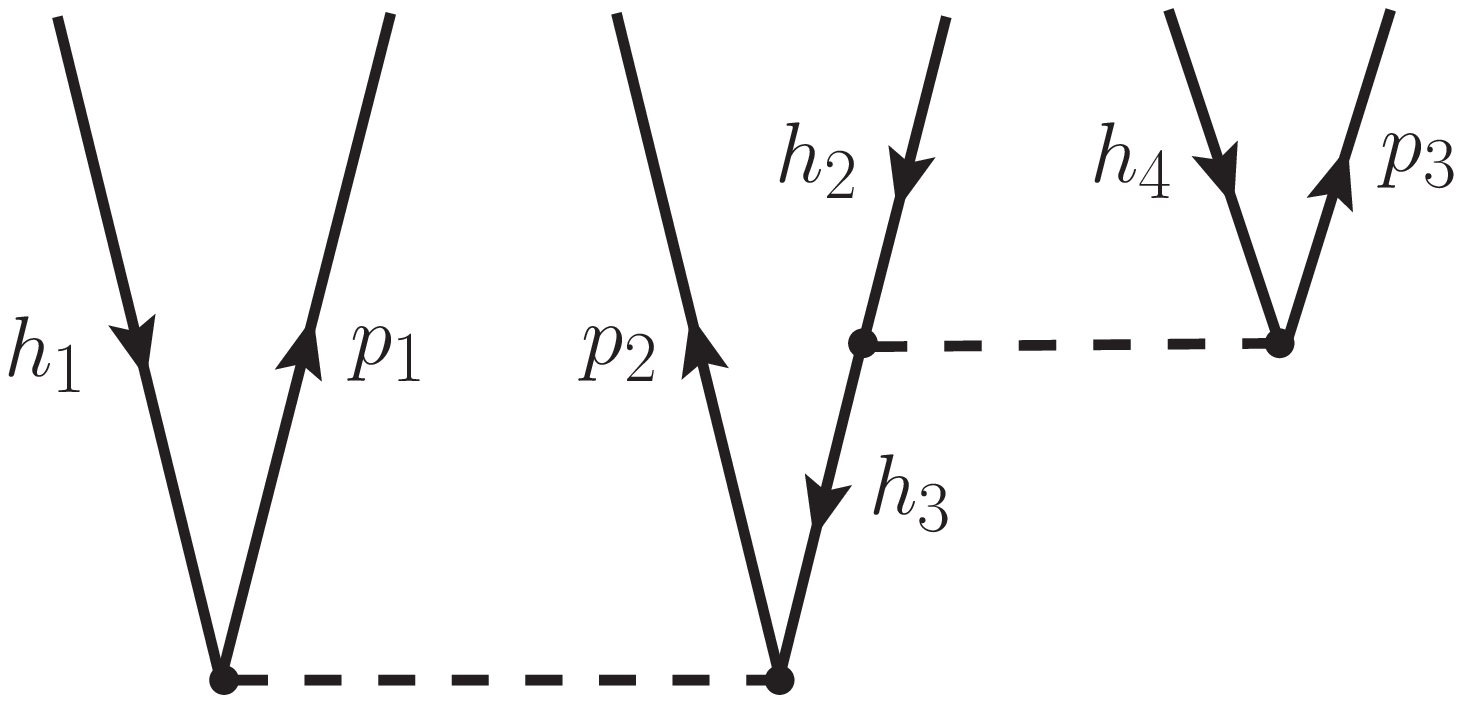}
}
\subfigure[]
{
\includegraphics[width=2.22in]{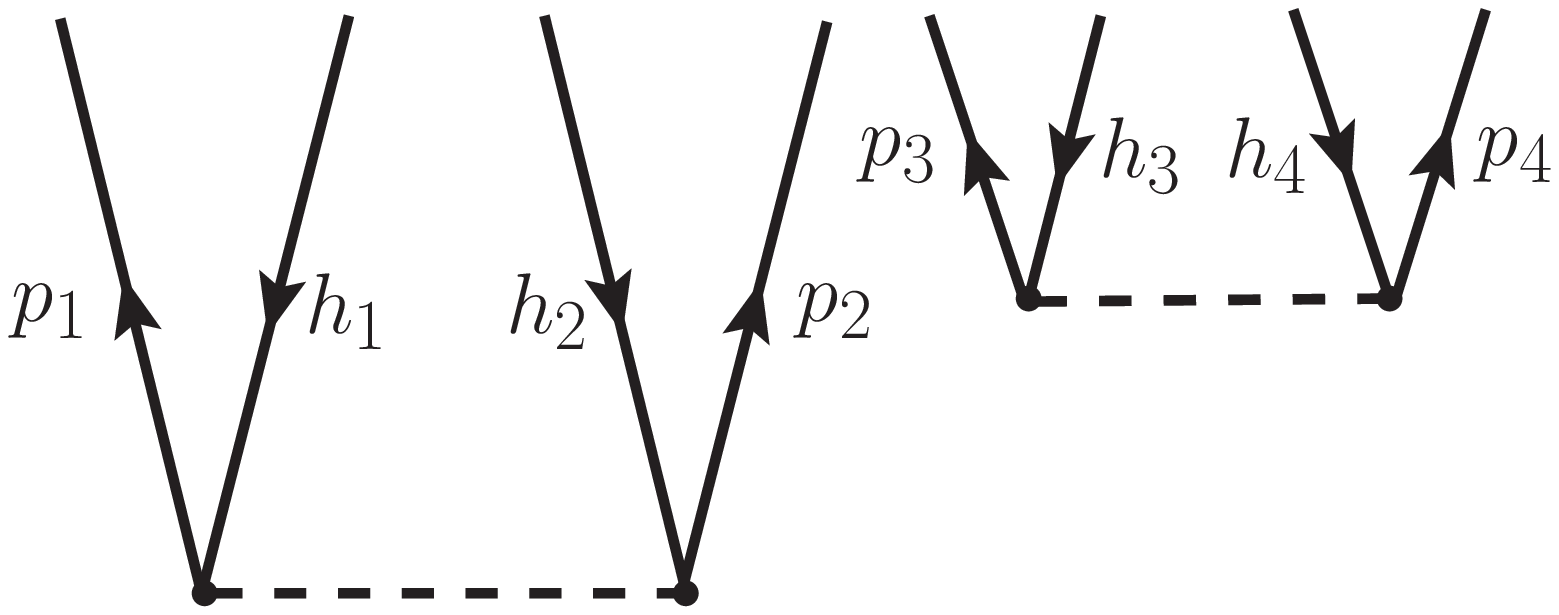}
}
}
}
\caption{ASG diagrams for the first- and second-order corrections to the wave function \cite{bartlett2009}.
The panel (a) is for the first-order correction,
while (b) (c) ... (i) are for the second-order correction.}
\label{fig:wave} 
\end{figure}
\end{widetext}

\begin{figure}
\setlength{\fboxrule}{0.7pt}
\fbox{
\shortstack[c]{
\subfigure[~$\rho_{a}$]
{
\includegraphics[width=1.5in]{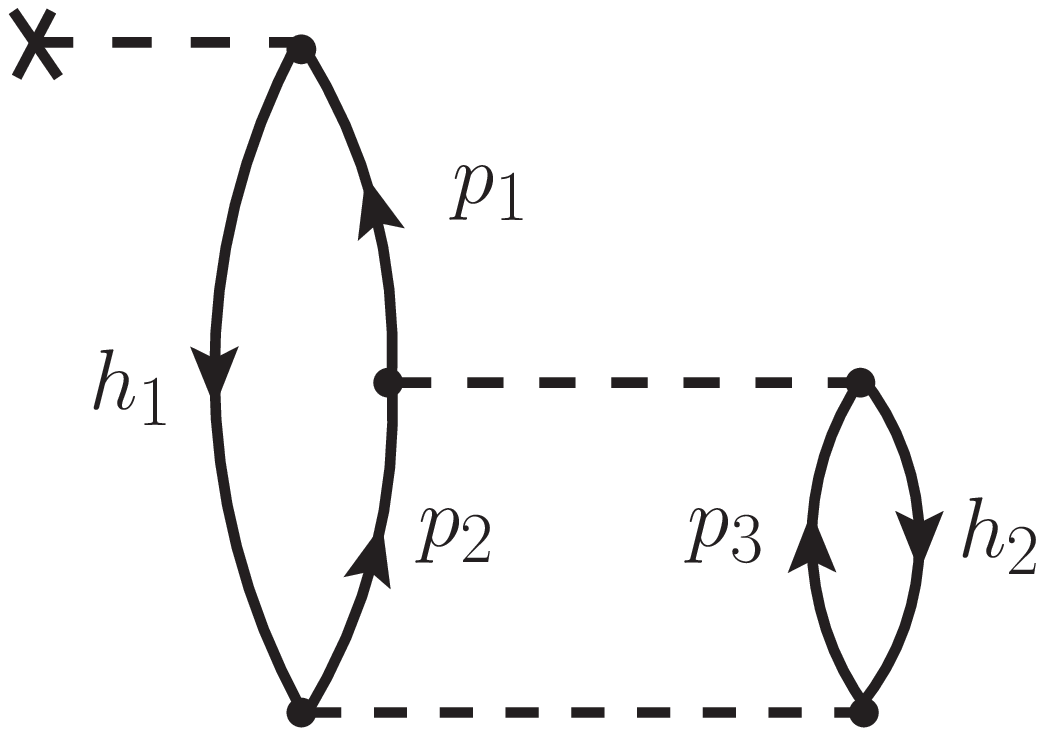}
}
\subfigure[~$\rho_{b}$]
{
\includegraphics[width=1.5in]{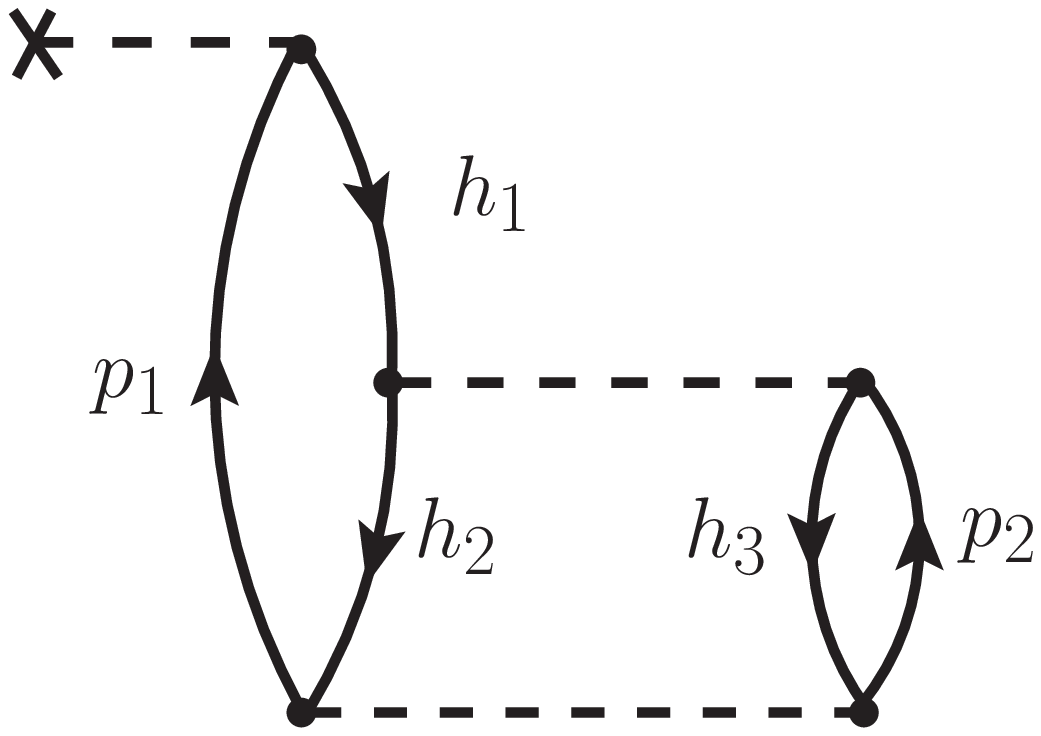}
}
\\
\subfigure[~$\rho_{c_{1}}$]
{
\includegraphics[width=1.5in]{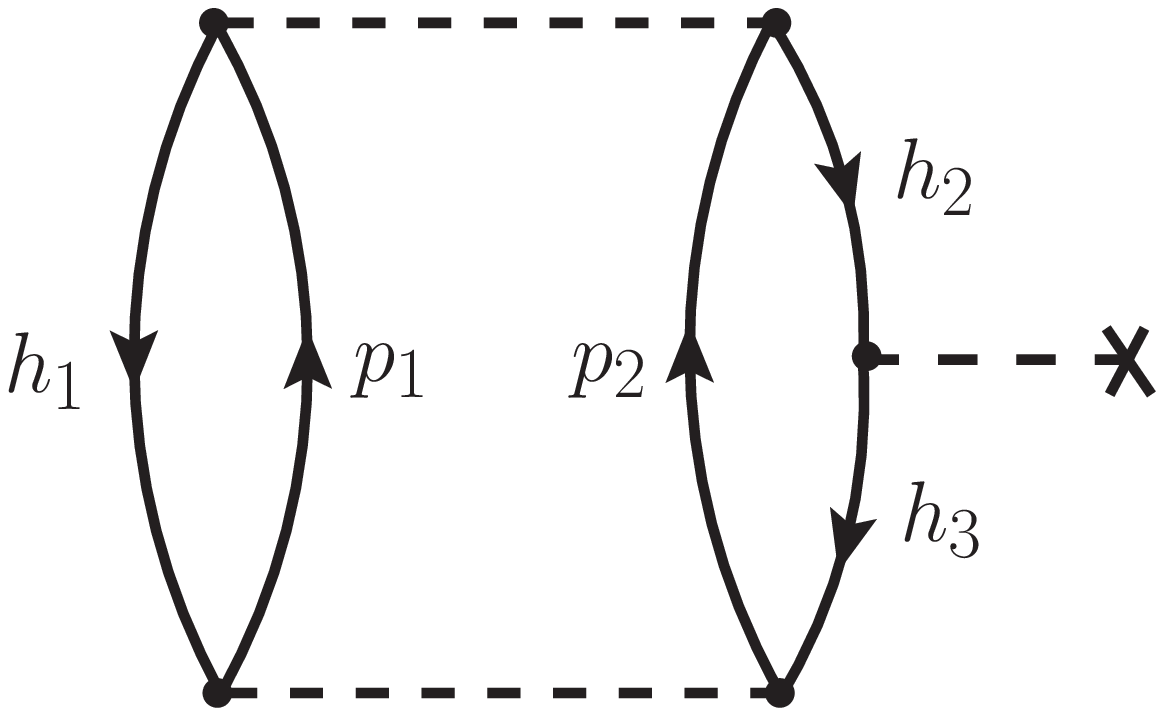}
}
\subfigure[~$\rho_{c_{2}}$]
{
\includegraphics[width=1.5in]{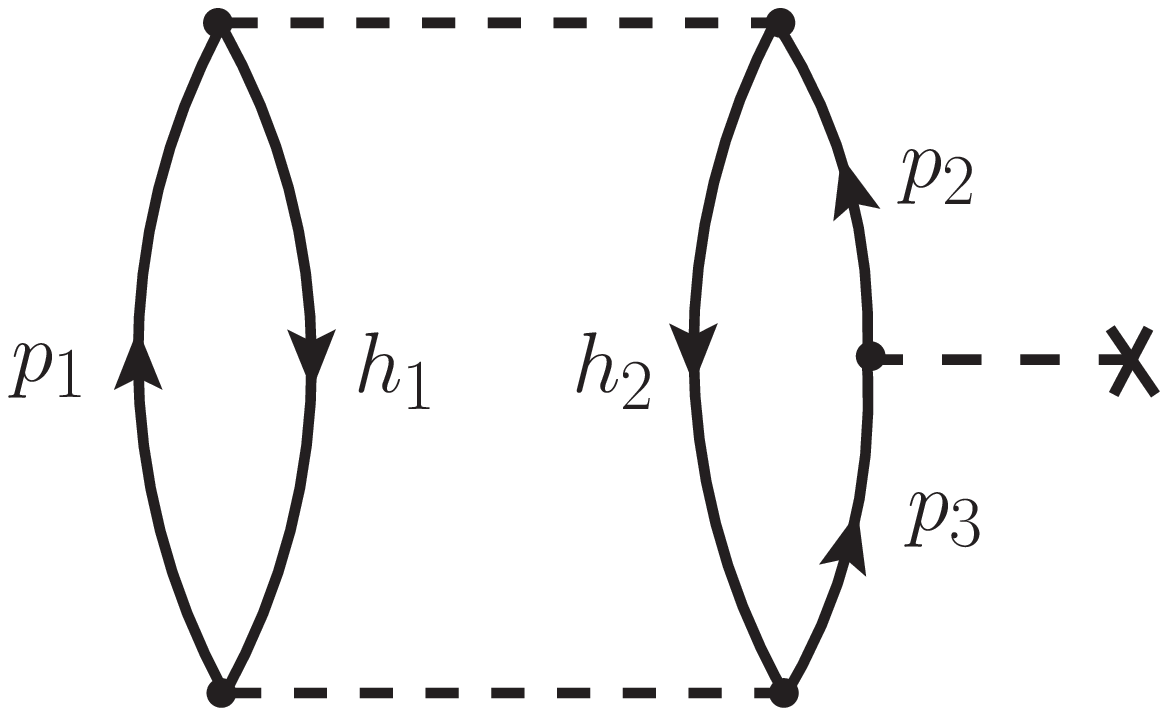}
}
}
}
\caption{ASG diagrams for the second-order corrections to the density.}
\label{fig:density} 
\end{figure}

The ASG diagrams for the first- and second-order corrections
to the wave function \cite{bartlett2009} are displayed
in  Fig.~\ref{fig:wave}.
The first-order wave function diagram, i.e., panel (a) in Fig.~\ref{fig:wave},
produces the second-order correction to the density.
While diagrams (b) and (c) of the second-order wave function correction
produce second-order corrections to the density,
other diagrams of the second-order wave function correction
contribute to higher-order corrections to the density.
The first- and second-order wave function corrections
which correct the density up to the second order can be written as
\begin{eqnarray}
\Psi^{(1)}= &&-\displaystyle\frac{1}{4}
\displaystyle\sum_{h_{1}h_{2}} \displaystyle\sum_{p_{1}p_{2}}
\frac{\langle p_{1} p_{2} |\text{\^{H}}| h_{1} h_{2}\rangle}
{(\varepsilon_{h_{1}}+\varepsilon_{h_{2}}-\varepsilon_{p_{1}}-\varepsilon_{p_{2}})}
\nonumber \\ &&\times
\left( c_{p_{1}}^{\dagger} c_{p_{2}}^{\dagger} c_{h_{2}} c_{h_{1}}| \Phi_{0}\rangle \right),
\end{eqnarray}
\begin{eqnarray}
\Psi^{(2)}_{b}= && \displaystyle\frac{1}{2}
\displaystyle\sum_{h_{1}h_{2}} \displaystyle\sum_{p_{1}p_{2}p_{3}}
\frac{\langle p_{1} h_{2} |\text{\^{H}}| p_{2} p_{3}\rangle
\langle p_{2} p_{3} |\text{\^{H}}| h_{1} h_{2}\rangle}
{(\varepsilon_{h_{1}}-\varepsilon_{p_{1}})
(\varepsilon_{h_{1}}+\varepsilon_{h_{2}}-\varepsilon_{p_{2}}-\varepsilon_{p_{3}})}
\nonumber \\ &&\times
\left(c_{p_{1}}^{\dagger} c_{h_{1}}| \Phi_{0}\rangle \right),
\end{eqnarray}
\begin{eqnarray}
\Psi^{(2)}_{c}= &&-\displaystyle\frac{1}{2}
\displaystyle\sum_{h_{1}h_{2}h_{3}} \displaystyle\sum_{p_{1}p_{2}}
\frac{\langle h_{2} h_{3} |\text{\^{H}}| h_{1} p_{2}\rangle
\langle p_{1} p_{2} |\text{\^{H}}| h_{2} h_{3}\rangle}
{(\varepsilon_{h_{1}}-\varepsilon_{p_{1}})
(\varepsilon_{h_{2}}+\varepsilon_{h_{3}}-\varepsilon_{p_{1}}-\varepsilon_{p_{2}})}
\nonumber \\ &&\times
\left(c_{p_{1}}^{\dagger} c_{h_{1}}| \Phi_{0}\rangle \right).
\end{eqnarray}
The total wave function that corrects the density up to the second order is
\begin{eqnarray}
\begin{array}{ll}
\Psi=\Phi_{0}+\Psi^{(1)}+\Psi^{(2)}_{b}+\Psi^{(2)}_{c}.
\end{array}
\end{eqnarray}
Then, the corrected density is written as
\begin{eqnarray}
\rho (\vec{r})&&=\langle \Psi| \hat{\rho} (\vec{r}) | \Psi \rangle
\nonumber \\ &&
=\langle \Phi_{0}| \hat{\rho} (\vec{r}) | \Phi_{0}\rangle+
\langle \Phi_{0}| \hat{\rho} (\vec{r}) | \Phi_{0}\rangle
\langle \Psi^{(1)}| \Psi^{(1)}\rangle
\nonumber \\ &&
\quad +2 \langle \Phi_{0}| \hat{\rho}_{N}| \Psi_{b}^{(2)}\rangle+
2 \langle \Phi_{0}|\hat{\rho}_{N}| \Psi_{c}^{(2)}\rangle+
\langle \Psi^{(1)}|\hat{\rho}_{N}| \Psi^{(1)}\rangle
\nonumber \\ &&
=\langle \Phi_{0}| \hat{\rho} (\vec{r}) | \Phi_{0}\rangle+
\langle \Phi_{0}| \hat{\rho} (\vec{r}) | \Phi_{0}\rangle
\langle \Psi^{(1)}| \Psi^{(1)}\rangle
\nonumber \\ &&
\quad +2 \rho_{a}+2 \rho_{b}+ \rho_{c_{1}}+ \rho_{c_{2}},
\end{eqnarray}
where $\rho_{a}=\langle \Phi_{0}| \hat{\rho}_{N}| \Psi_{b}^{(2)}\rangle$,
$\rho_{b}=\langle \Phi_{0}|\hat{\rho}_{N}| \Psi_{c}^{(2)}\rangle$ and
$\rho_{c_{1}}+\rho_{c_{2}}=\langle \Psi^{(1)}|\hat{\rho}_{N}| \Psi^{(1)}\rangle$.
They are displayed using the language of the diagram in Fig.~\ref{fig:density}.
Dashed lines with cross contribute to the reduced matrix elements
 $\langle \nu_{1} l j\| \rho \| \nu_{2} l j\rangle=
 \sqrt{2j+1}\langle \nu_{1} l j m_{j} | \rho | \nu_{2} l j m_{j}\rangle$.

The detailed formulae of the density correction terms in the angular momentum coupled scheme
are written as
\begin{eqnarray}
\rho_{a}=&& \frac{1}{2}\displaystyle\sum_{h_{1},h_{2}}
\displaystyle\sum_{p_{1},p_{2},p_{3}}
\frac{(-1)^{j_{h_{1}}+j_{h_{2}}}\sqrt{2j_{h_{2}}+1}}
{(\varepsilon_{h_{1}}-\varepsilon_{p_{1}})
(\varepsilon_{h_{1}}+\varepsilon_{h_{2}}-\varepsilon_{p_{2}}-\varepsilon_{p_{3}})}
\nonumber \\ \times &&
\displaystyle\sum_{J}(-1)^{J}(2J+1)
\left\{
  \begin{array}{ccc}
    j_{h_{1}} & j_{p_{1}} &  0\\
    j_{h_{2}} & j_{h_{2}} &  J
  \end{array}
\right\}
\langle (h_{1} h_{2})J |\text{\^{H}}|(p_{2} p_{3})J\rangle
\nonumber \\ \times &&
\langle (p_{2} p_{3})J |\text{\^{H}}|(p_{1} h_{2})J\rangle
\langle h_{1} \| \rho \| p_{1}\rangle,
\end{eqnarray}
\begin{eqnarray}
\rho_{b}=&& -\frac{1}{2}\displaystyle\sum_{h_{1},h_{2},h_{3}}
\displaystyle\sum_{p_{1},p_{2}}
\frac{(-1)^{j_{h_{1}}+j_{p_{2}}}\sqrt{2j_{p_{2}}+1}}
{(\varepsilon_{h_{1}}-\varepsilon_{p_{1}})
(\varepsilon_{h_{2}}+\varepsilon_{h_{3}}-\varepsilon_{p_{1}}-\varepsilon_{p_{2}})}
\nonumber \\ \times &&
\displaystyle\sum_{J}(-1)^{J}(2J+1)
\left\{
  \begin{array}{ccc}
    j_{h_{1}} & j_{p_{1}} &  0\\
    j_{p_{2}} & j_{p_{2}} &  J
  \end{array}
\right\}
\langle (p_{1} p_{2})J |\text{\^{H}}|(h_{2} h_{3})J\rangle
\nonumber \\ \times &&
\langle (h_{2} h_{3})J |\text{\^{H}}|(h_{1} p_{2})J\rangle
\langle h_{1} \| \rho \| p_{1}\rangle,
\end{eqnarray}
\begin{eqnarray}
\rho_{c_{1}}=&& -\frac{1}{2}\displaystyle\sum_{h_{1},h_{2},h_{3}}
\displaystyle\sum_{p_{1},p_{2}}
\frac{(-1)^{j_{h_{1}}+j_{h_{2}}}\sqrt{2j_{h_{1}}+1}}
{(\varepsilon_{h_{1}}+\varepsilon_{h_{2}}-\varepsilon_{p_{1}}-\varepsilon_{p_{2}})
(\varepsilon_{h_{1}}+\varepsilon_{h_{3}}-\varepsilon_{p_{1}}-\varepsilon_{p_{2}})}
\nonumber \\ \times &&
\displaystyle\sum_{J}(-1)^{J}(2J+1)
\left\{
  \begin{array}{ccc}
    j_{h_{1}} & j_{h_{1}} &  0\\
    j_{h_{2}} & j_{h_{3}} &  J
  \end{array}
\right\}
\langle (h_{1} h_{2})J |\text{\^{H}}|(p_{1} p_{2})J\rangle
\nonumber \\ \times &&
\langle (p_{1} p_{2})J |\text{\^{H}}|(h_{1} h_{3})J\rangle
\langle h_{3} \| \rho \| h_{2}\rangle,
\end{eqnarray}
\begin{eqnarray}
\rho_{c_{2}}=&& \frac{1}{2}\displaystyle\sum_{h_{1},h_{2}}
\displaystyle\sum_{p_{1},p_{2},p_{3} }
\frac{(-1)^{j_{p_{1}}+j_{p_{3}}}\sqrt{2j_{p_{1}}+1}}
{(\varepsilon_{h_{1}}+\varepsilon_{h_{2}}-\varepsilon_{p_{1}}-\varepsilon_{p_{3}})
(\varepsilon_{h_{1}}+\varepsilon_{h_{2}}-\varepsilon_{p_{1}}-\varepsilon_{p_{2}})}
\nonumber \\ \times &&
\displaystyle\sum_{J}(-1)^{J}(2J+1)
\left\{
  \begin{array}{ccc}
    j_{p_{1}} & j_{p_{1}} &  0\\
    j_{p_{3}} & j_{p_{2}} &  J
  \end{array}
\right\}
\langle (p_{1} p_{3})J |\text{\^{H}}|(h_{1} h_{2})J \rangle
\nonumber \\ \times &&
\langle (h_{1} h_{2})J |\text{\^{H}}|(p_{1} p_{2})J\rangle
\langle p_{2} \| \rho \| p_{3}\rangle,
\end{eqnarray}
where $\left\{
  \begin{array}{ccc}
    j_{1} & j_{2} &  j_{3}\\
    j_{4} & j_{5} & j_{6} \\
  \end{array}
\right\}$ is Wigner 6-j symbol.
The letters $h_{1},h_{2},...$
indicate occupied single-particle levels in $|\text{HF}\rangle$
(i.e., hole states),
the letters $p_{1},p_{2},...$ for unoccupied levels (i.e., particle states).
$\varepsilon_{h}$ or $\varepsilon_{p}$ is the energy of particle or hole state, respectively.
States $h$ or $p$
includes the quantum numbers of the orbital angular momentum $l$,
total angular momentum $j$,
isospin projection quantum number $m_{t}$,
and additional quantum number $\nu$,
i.e., $|h\rangle$ or $|p\rangle=|\nu ljt_{z}\rangle$.
We define an anti-symmetrized two-particle state (unnormalized) coupled
to a good angular momentum $J$ with a projection $M$,
\begin{eqnarray}
| (j_{1} j_{2})J M\rangle=\displaystyle\sum_{m_{1},m_{2}}
\langle j_{1} m_{1}j_{2} m_{2}| J M \rangle
| (j_{1} m_{1}) (j_{2} m_{2})\rangle.
\end{eqnarray}

\subsubsection{ Root-mean-square radii}

The root-mean-square (rms) radius
is an important  global indicator for the change of the density distribution
arising from correlations beyond HF.
The squares of the rms radii for point-like
 proton, neutron and nucleon (matter) distributions
 are the averaged values of the operators \cite{PhysRevC.56.191}, respectively,
\begin{eqnarray}
\begin{array}{ll}
{\hat{r}_{\text{pp}}}^{2}=\displaystyle\frac{1}{Z} \displaystyle\sum_{i=1}^{Z} (\vec{r_{i}}-\vec{r_{0}})^{2},
\end{array}
\label{r2-1}
\end{eqnarray}
\begin{eqnarray}
\begin{array}{ll}
{\hat{r}_{\text{nn}}}^{2}=\displaystyle\frac{1}{N} \displaystyle\sum_{i=1}^{N} (\vec{r_{i}}-\vec{r_{0}})^{2},
\end{array}
\label{r2-2}
\end{eqnarray}
\begin{eqnarray}
\begin{array}{ll}
{\hat{r}_{\text{m}}}^{2}=\displaystyle\frac{1}{A} \displaystyle\sum_{i=1}^{A} (\vec{r_{i}}-\vec{r_{0}})^{2}
           = \frac{1}{A^{2}} \displaystyle\sum_{i<j}^{A} (\vec{r_{i}}-\vec{r_{j}})^{2},
\end{array}
\label{r2-3}
\end{eqnarray}
with the c.m. position $\vec{r_{0}}=\frac{1}{A} \displaystyle\sum_{i=1}^{A}\vec{r_{i}}$.
The charge radius $r_{\text{ch}}$ obtained
from the point-proton radius $r_{\text{pp}}$
using the standard expression
\cite{PhysRevC.91.051301}
\begin{eqnarray}
\label{charge}
\langle r_{\text{ch}}^{2}\rangle =\langle r_{\text{pp}}^{2}\rangle +
 R_{\text{p}}^{2} + \dfrac{N}{Z} R_{\text{n}}^{2}
+\dfrac{3 \hbar^{2}}{4m_{\text{p}}^{2}c^{2}},
\end{eqnarray}
where $\dfrac{3 \hbar^{2}}{4m_{\text{p}}^{2}c^{2}} \approx 0.033 \; \text{fm}^{2}$, $R_{\text{n}}^{2}=-0.1149(27)\; \text{fm}^{2}$,
$R_{\text{p}}=0.8775(51)$ fm.
The point-proton or point-neutron rms radius operator is a two-body operator.
The squares of the rms radii can be calculated either from the translational invariant local density
or directly using the two-body operators [ i.e.,
Eqs.~(\ref{r2-1}), (\ref{r2-2}) and (\ref{r2-3}) ].
Since we adopt MBPT with intermediate normalization
[ i.e., Eqs.~(\ref{norm}) ],
the perturbed wave function is unnormalized.
In the present work,
we use the one-body local density to calculate the radius, as
\begin{eqnarray}
\label{Rpp}
\langle{R^{2}_{\text{pp}}}\rangle=\displaystyle\frac{\int r^{2} \rho_{\text{p}}(\vec{r}) d^{3}r }{\int \rho_{\text{p}}(\vec{r}) d^{3}r}.
\end{eqnarray}

The wave function is written in the laboratory HO coordinate,
starting from an anti-symmetrized Slater determinant
which contains the component of the center-of-mass (c.m.) motion.
Consequently, the local one-body density calculated with
the wave function includes contribution from the c.m. motion.
The c.m. correction to the radius can be approximated as follows.
Eq.~(\ref{r2-3}) gives
\begin{eqnarray}
\begin{array}{ll}
{\hat{r}_{\text{m}}}^{2}= \displaystyle\frac{1}{A^{2}} \displaystyle\sum_{i<j}^{A} (\vec{r_{i}}-\vec{r_{j}})^{2}
=\left( 1-\frac{1}{A} \right)\cdot \left({\displaystyle\sum_{i=1}^{A}\vec{r_{i}}^{2}}/{A} \right)
-\frac{2}{A^{2}}\cdot \left( \displaystyle\sum_{i<j}^{A} \vec{r_{i}} \cdot \vec{r_{j}} \right).
\end{array}
\end{eqnarray}
If the cross term $ \displaystyle\sum_{i<j}^{A} \vec{r_{i}} \cdot \vec{r_{j}} $ is neglected,
we have
\begin{eqnarray}
\begin{array}{ll}
{\hat{r}_{\text{m}}}^{2}\approx \left( 1-\displaystyle\frac{1}{A} \right)\cdot \left({\displaystyle\sum_{i=1}^{A}\vec{r_{i}}^{2}}/{A} \right)
\end{array}.
\end{eqnarray}
Similarly for the proton radius,
\begin{eqnarray}
\begin{array}{ll}
{\hat{r}_{\text{pp}}}^{2}\approx \left( 1-\displaystyle\frac{1}{A} \right)\cdot \left({\displaystyle\sum_{i=1}^{Z}\vec{r_{i}}^{2}}/{Z} \right)
\end{array}.
\end{eqnarray}
This gives an approximate c.m. correction to the point-proton rms radius,
\begin{eqnarray}
\label{r-com}
\begin{array}{ll}
\Delta r_{\text{c.m.}}=\left[\left(1- \displaystyle\frac{1}{A}\right)\cdot \langle R_{\text{pp}}^{2} \rangle \right]^{1/2}
- \langle R_{\text{pp}}^{2} \rangle ^{1/2},
\end{array}
\end{eqnarray}
where $\langle R_{\text{pp}}^{2} \rangle ^{1/2}$
is the point-proton rms radius calculated by Eq~(\ref{Rpp}).
Then the rms radius of the point-proton distribution is obtained by
\begin{eqnarray}
r_{\text{pp}}=\langle R_{\text{pp}}^{2} \rangle ^{1/2}+\Delta r_{\text{c.m.}}.
\end{eqnarray}

\section{Calculations and discussions}

In this section, we apply the method outlined in Section~\ref{sec:1}
to two light closed-shell nuclei, $^{4}$He and $^{16}$O.
The SRG-softened chiral N$^{3}$LO and the ``bare'' JISP16 interactions
are adopted for the effective Hamiltonians.

\subsection{Calculations with chiral N$^3$LO interaction}

The SHF is carried out within the HO basis.
The HO basis is truncated by a cutoff according to
the number $N_{\rm shell} = \text{max}(2n+l +1)$,
where $N_{\rm shell}$ indicates how many major HO shells are included in the truncation.
After the SHF calculation,
the MBPT corrections are calculated in the SHF basis.
In the present calculations,
the basis spaces employed take $N_{\rm shell}$=7, 9, 11 and 13.
We verify that such a truncation is sufficient
for the converged calculations of the ground state energies for these magic nuclei $^4$He and $^{16}$O.
\begin{figure}
\includegraphics[scale=0.50]{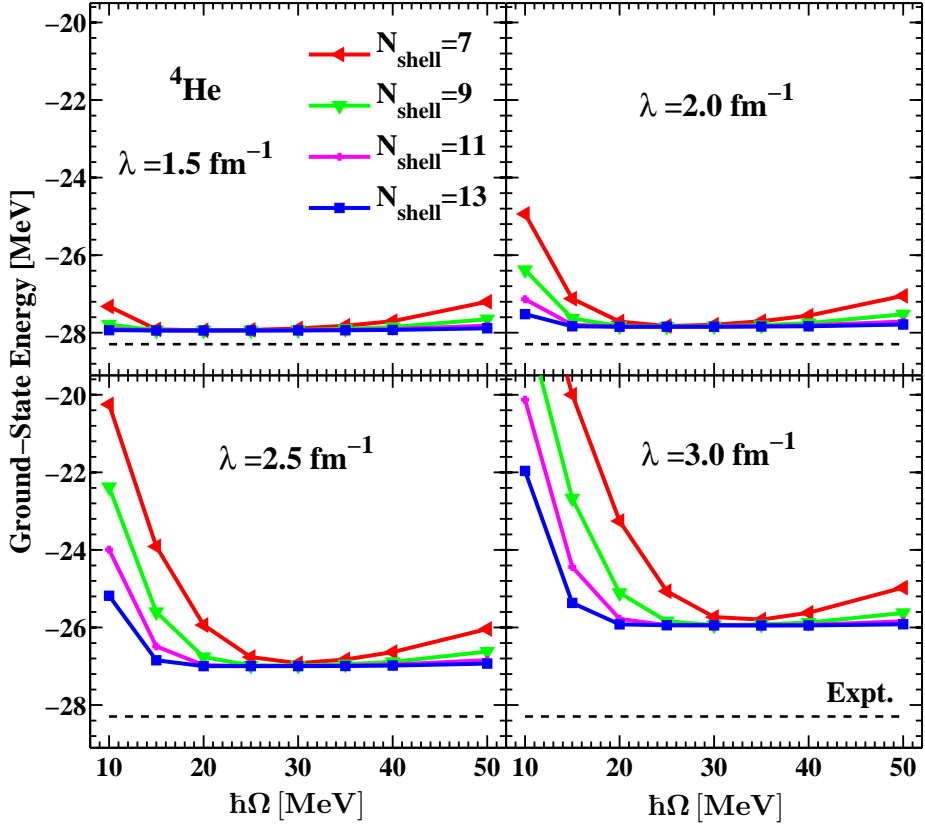}
\caption{\label{fig:he4} HF-MBPT calculations of $^{4}\text{He}$
ground-state energy through third order as a function of oscillator parameter $\hbar \Omega$
with the chiral N$^{3}$LO potential \cite{PhysRevC.68.041001,Machleidt20111}
renormalized by SRG at different softening parameters
$\lambda=1.5, 2.0, 2.5, 3.0$ $\text{fm}^{-1}$.
The dashed line represents the experimental ground-state energy.}
\end{figure}
\begin{figure}
\setlength{\abovecaptionskip}{0pt}
\setlength{\belowcaptionskip}{0pt}
\includegraphics[scale=0.50]{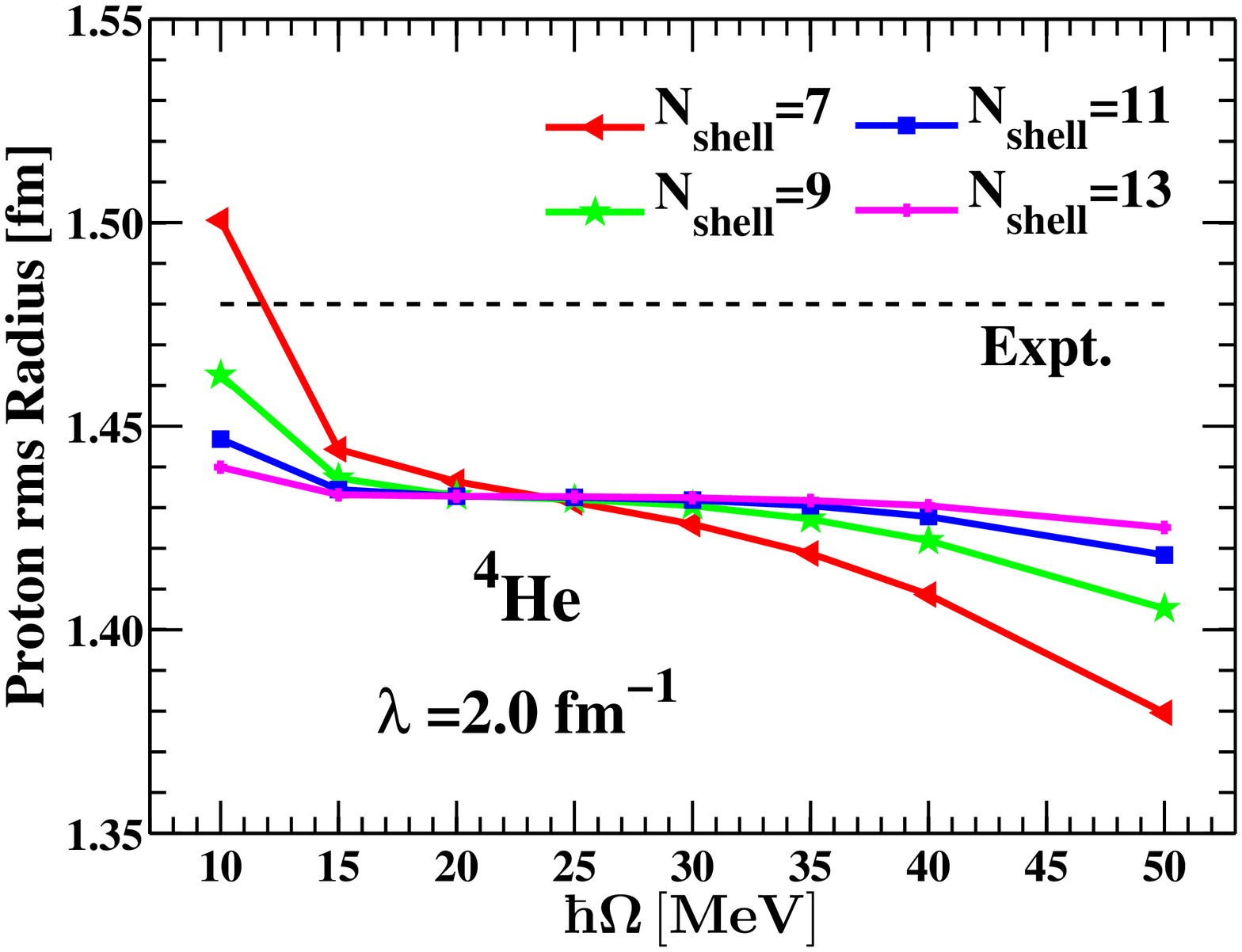}
\caption{\label{fig:he4_n3lo_rms}
Point-proton rms radius of $^{4}$He as a function of oscillator parameter $\hbar \Omega$
with different $N_{\text{shell}}$.
The chiral N$^{3}$LO potential \cite{PhysRevC.68.041001,Machleidt20111} is softened by the SRG method.}
\end{figure}
\begin{table}
\caption{
\label{tab:n3lo-he4-e}
Ground-state energy (in MeV) of $^4$He, analyzed in order-by-order HF-MBPT calculations with N$^3$LO softened at different SRG-softening parameter values ($\lambda$). PT2 and PT3 represent the second- and third-order corrections to energy, respectively. We take $N_{\text{shell}}=13$ and $\hbar\Omega=35$ MeV.
}
\begin{ruledtabular}
\begin{tabular}{ccccc p{cm}}
\multicolumn{1}{c}{} & \multicolumn{4}{c}{SRG flow parameter $\lambda$ (fm$^{-1}$)} \\
\cline{2-5}
\textrm{} & \textrm{1.5} &\textrm{2.0}& \textrm{2.5}& \textrm{3.0}\\
\colrule
 Expt.~\cite{1674-1137-36-12-001}                   & -28.296 & -28.296 & -28.296 & -28.296\\
 NCSM ~\cite{PhysRevC.87.054312}                   & -28.20  & -28.41  & -27.43  & -26.80 \\
  SHF                    & -25.754 & -21.864 & -15.854 & -10.278\\
 PT2                     &  -1.788 &  -5.088 &  -9.652 & -13.783\\
 PT3                     &  -0.391 &  -0.899 &  -1.523 &  -1.953\\
 SHF+PT2+PT3             & -27.933 & -27.850 & -27.029 & -26.013\\
\end{tabular}
\end{ruledtabular}
\end{table}
\begin{table}
\caption{
\label{tab:n3lo-he4-r}
Point-proton rms radius (in fm) of $^4$He in the HF-MBPT calculations with N$^3$LO softened at different SRG-softening parameter values. PT2 designates the second-order correction to the radius. $N_{\text{shell}}=13$ and $\hbar\Omega=35$ MeV are taken. The experimental point-proton rms radius is obtained using Eq.~(\ref{charge}) with the experimental charge radius taken from \cite{Angeli201369}.
}
\begin{ruledtabular}
\begin{tabular}{ccccc p{cm}}
\multicolumn{1}{c}{} & \multicolumn{4}{c}{SRG flow parameter $\lambda$ (fm$^{-1}$)} \\
\cline{2-5}
\textrm{} & \textrm{1.5} &\textrm{2.0}& \textrm{2.5}& \textrm{3.0}\\
\colrule
 Expt.                           &  1.477 & 1.477 & 1.477 & 1.477\\
  SHF                            &  1.677 & 1.652 & 1.714 & 1.816\\
 PT2                             &  0.007 & 0.001 &-0.021 &-0.065\\
 $\Delta r_{\text{c.m.}}$        & -0.226 & -0.222&-0.227 &-0.235\\
 SHF+PT2+$\Delta r_{\text{c.m.}}$&  1.458 & 1.431 & 1.466 & 1.516\\
\end{tabular}
\end{ruledtabular}
\end{table}

Fig.~\ref{fig:he4} shows the MBPT calculated ground-state energy of $^4$He.
The calculations were done with the chiral N$^3$LO interaction
which was renormalized by SRG.
We see that good convergence of the calculated energy
by virtue of independence from
the oscillator parameter $\hbar\Omega$
and $N_{\text{shell}}$ is obtained at least for the truncations $N_{\text{shell}}=11$ and 13.
We note that the dependence on the parameter $\hbar\Omega$
displays behavior similar to NCSM calculations \cite{Bogner200821,PhysRevC.87.054312}.
The softening parameter $\lambda=3.0$ fm$^{-1}$ seems to be insufficient to produce an interaction soft enough for good convergence in MBPT.
Jurgenson {\it et al.}, have investigated
the SRG evolution with the softening parameter $\lambda$ in $^4$He at $\hbar\Omega=36$ MeV \cite{PhysRevLett.103.082501,PhysRevC.87.054312}.
They found that $\lambda \approx 2.0$ fm$^{-1}$
can reasonably reproduce the experimental $^4$He ground-state energy
with the $NN$-only interaction (without requiring a three-body force).

Fig.~\ref{fig:he4_n3lo_rms} shows the radius calculations at different $\hbar\Omega$ with $\lambda=2.0$ fm$^{-1}$.
Tables~\ref{tab:n3lo-he4-e} and ~\ref{tab:n3lo-he4-r} give the details of the HF-MBPT calculations
with  different $\lambda$ values.
We see that both second- and third-order corrections to energy decrease with decreasing $\lambda$.
This is easily understood because MBPT mainly treats intermediate-range correlations and
these correlations are weakened with decreasing $\lambda$.
With sufficiently small $\lambda$, higher-order corrections to the energy can be neglected.
The second-order correction to the radius is already small, which decreases with decreasing $\lambda$ in $^4$He. The c.m. correction to the radius is larger than the MBPT correction. It may be concluded that, at least for $^4$He, MBPT corrections up to third order in energy and up to second order in radius within the HF basis should give converged results
for $\lambda$ below about 3.0 fm$^{-1}$.
It has been pointed out that the MBPT calculation
within the HO basis could be divergent even for softened interactions \cite{Tichai:2016joa}.
The Hamiltonian (1) is written already in the relative coordinate,
and SHF can preserve the translational invariance for the ground state energy \cite{epja14413}
so that no c.m. correction is needed for the ground state energy.
\begin{figure}
\setlength{\abovecaptionskip}{10pt}
\setlength{\belowcaptionskip}{10pt}
\includegraphics[scale=0.50]{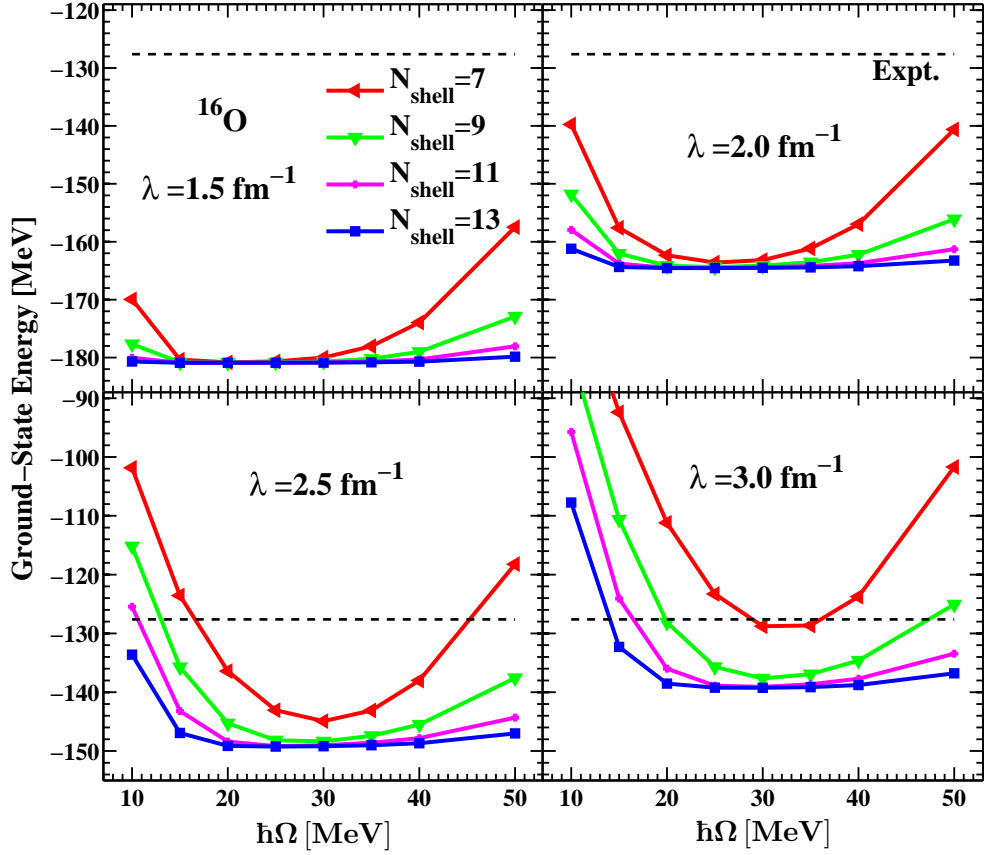}
\caption{\label{fig:o16}
HF-MBPT calculations of $^{16}\text{O}$ as a function of oscillator parameter $\hbar \Omega$
with the chiral N$^{3}$LO potential \cite{PhysRevC.68.041001,Machleidt20111}
renormalized by SRG at different softening parameters
$\lambda=1.5, 2.0, 2.5, 3.0$ $\text{fm}^{-1}$.}
\end{figure}
\begin{figure}
\setlength{\abovecaptionskip}{0pt}
\setlength{\belowcaptionskip}{0pt}
\includegraphics[scale=0.50]{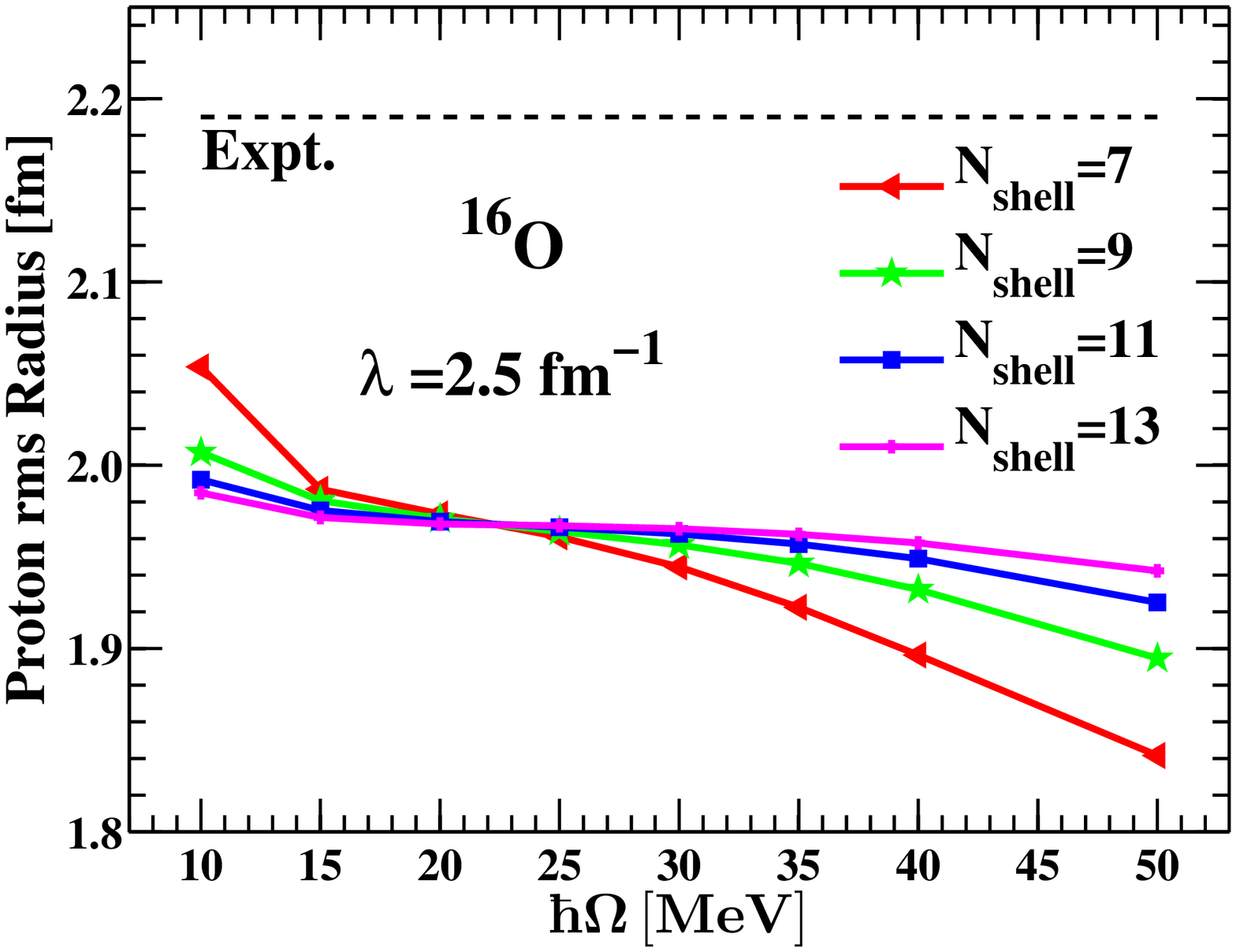}
\caption{\label{fig:o16_n3lo_rms}
Point-proton rms radius of $^{16}$O as a function of oscillator parameter $\hbar \Omega$
with different $N_{\text{shell}}$.
The chiral N$^{3}$LO potential \cite{PhysRevC.68.041001,Machleidt20111} is softened by the SRG method.}
\end{figure}
\begin{table}
\caption{
\label{tab:n3lo-o16-e}
Ground-state energy (in MeV) of $^{16}$O, analyzed in order-by-order HF-MBPT calculations with N$^3$LO softened at different SRG-softening parameter values ($\lambda$). We take $N_{\text{shell}}=13$ and $\hbar\Omega=35$ MeV.
}
\begin{ruledtabular}
\begin{tabular}{ccccc p{cm}}
\multicolumn{1}{c}{} & \multicolumn{4}{c}{SRG flow parameter $\lambda$ (fm$^{-1}$)} \\
\cline{2-5}
\textrm{} & \textrm{1.5} &\textrm{2.0}& \textrm{2.5}& \textrm{3.0}\\
\colrule
 Expt.~\cite{1674-1137-36-12-001}                   & -127.619 & -127.619 & -127.619 & -127.619\\
  SHF                    & -169.968 & -133.169 &  -85.173 &  -44.102\\
 PT2                     &  -10.132 &  -29.497 &  -59.617 &  -88.326\\
 PT3                     &   -0.794 &   -1.931 &   -4.630 &   -7.339\\
 SHF+PT2+PT3             & -180.893 & -164.597 & -149.419 & -139.767\\
\end{tabular}
\end{ruledtabular}
\end{table}
\begin{table}
\caption{
\label{tab:n3lo-o16-r}
Point-proton rms radius (in fm) of $^{16}$O in the HF-MBPT calculations with N$^3$LO softened at different SRG-softening parameter values. $N_{\text{shell}}=13$ and $\hbar\Omega=35$ MeV are taken. The experimental point-proton rms radius is obtained using Eq.~(\ref{charge}) with the experimental charge radius taken from \cite{Angeli201369}.
}
\begin{ruledtabular}
\begin{tabular}{ccccc p{cm}}
\multicolumn{1}{c}{} & \multicolumn{4}{c}{SRG flow parameter $\lambda$ (fm$^{-1}$)} \\
\cline{2-5}
\textrm{} & \textrm{1.5} &\textrm{2.0}& \textrm{2.5}& \textrm{3.0}\\
\colrule
 Expt.                           &  2.581 & 2.581 & 2.581 & 2.581\\
  SHF                            &  2.098 & 2.096 & 2.201 & 2.345\\
 PT2                             &  0.011 & 0.011 &-0.006 &-0.042\\
 $\Delta r_{\text{c.m.}}$        & -0.067 &-0.067 &-0.070 &-0.073\\
 SHF+PT2+$\Delta r_{\text{c.m.}}$&  2.042 & 2.040 & 2.125 & 2.230\\
\end{tabular}
\end{ruledtabular}
\end{table}

Fig.~\ref{fig:o16} shows the energy calculations for $^{16}$O.
The convergence behavior is similar to that in $^4$He.
The $N_{\rm shell}=11$ and 13 calculations appear nearly convergent.
However, calculations with small $\lambda$ values (e.g., $\leq 2.0$ fm$^{-1}$)
give over-binding, compared with data.
This phenomenon should be more obvious for heavier nuclei.
The main reason is that the three-body
and higher-order forces are omitted in these calculations.
The emergence of induced three-body forces
and beyond is related to the SRG softening parameter $\lambda$.
A larger $\lambda$ value evolves a harder effective $NN$ potential.
In large $\lambda$ cases (e.g., $\lambda>3.0$ fm$^{-1}$),
effects from induced three-body and higher-order forces are small.
But a large $\lambda$ value may not sufficiently soften
the short-range correlations of the realistic force,
leading to demands for an excessively large model space
and increased dependence on higher-order corrections.
While a small $\lambda$ value may sufficiently soften the potential,
the contribution from induced three-body force may be not ignorable.
Within SRG, $\lambda\sim 2.0-2.5$ fm$^{-1}$ seems to be
an optimal range in which the $NN$ interaction
can be softened reasonably
and the combined three-body (initial plus induced) effects are greatly reduced \cite{PhysRevC.87.054312,PhysRevC.75.061001,Tichai:2016joa}.

The calculation of the radius for $^{16}$O is displayed in Fig.~\ref{fig:o16_n3lo_rms}.
Reasonable convergence is obtained for $N_{\text{shell}}=11$ and 13.
But the calculated radius is smaller than the experimental value.
It seems that other {\it ab initio} results yield radii that are
systematically smaller than experiment \cite{PhysRevC.73.044312,PhysRevC.91.051301}.
In Tables~\ref{tab:n3lo-o16-e} and ~\ref{tab:n3lo-o16-r},
 we give the order-by-order results of the HF-MBPT $^{16}$O calculations with the same parameters as those in $^4$He (i.e., $N_{\text{shell}}=13$ and $\hbar\Omega=35$ MeV) at different $\lambda$ values. The situation is similar to that in $^4$He. We can see that smaller contributions from the neglected higher-order corrections decrease with decreasing $\lambda$, and good convergence is obtained for the MBPT calculations within the HF basis at small $\lambda$ values.
It has pointed out that in the HF basis the fourth- and higher-order MBPT corrections are known
to be negligible in some cases \cite{Tichai:2016joa}.


\subsection{Calculations with the ``bare'' JISP16 potential}

As mentioned in the Introduction, the JISP16 interaction is established by the $J$-matrix technique,
and its parameters were determined by fitting both $NN$ scattering data
and nuclear structure data up to $A=16$ \cite{Shirokov200733}.
It is called ``bare'' because we, along with others,
do not apply renormalization procedures in order to use it in nuclear structure calculations.
To fit selected nuclear properties, the interaction has been tuned with phase-equivalent transformations to minimize the role of neglected many-body interactions.
This tuning exploits the residual freedoms in the off-shell properties of the NN interaction \cite{latePolyzou}.

\begin{figure}
\setlength{\abovecaptionskip}{1pt}
\setlength{\belowcaptionskip}{1pt}
\includegraphics[scale=0.40]{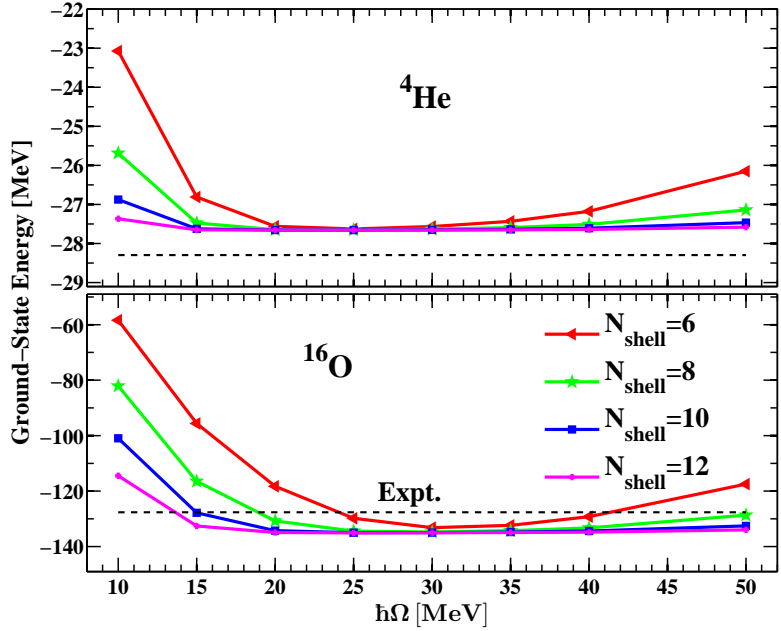}
\caption{\label{fig:jispo16} Ground-state binding energies of $^{4}$He and $^{16}$O as a function of the oscillator parameter $\hbar \Omega$
for different $N_{\text{shell}}$.
The ``bare'' JISP16 potential \cite{PhysRevC.70.044005,Shirokov200596,Shirokov200733} is used. The dashed lines represent the experimental ground state energies.}
\end{figure}
\begin{figure}
\setlength{\abovecaptionskip}{0pt}
\setlength{\belowcaptionskip}{0pt}
\includegraphics[scale=0.50]{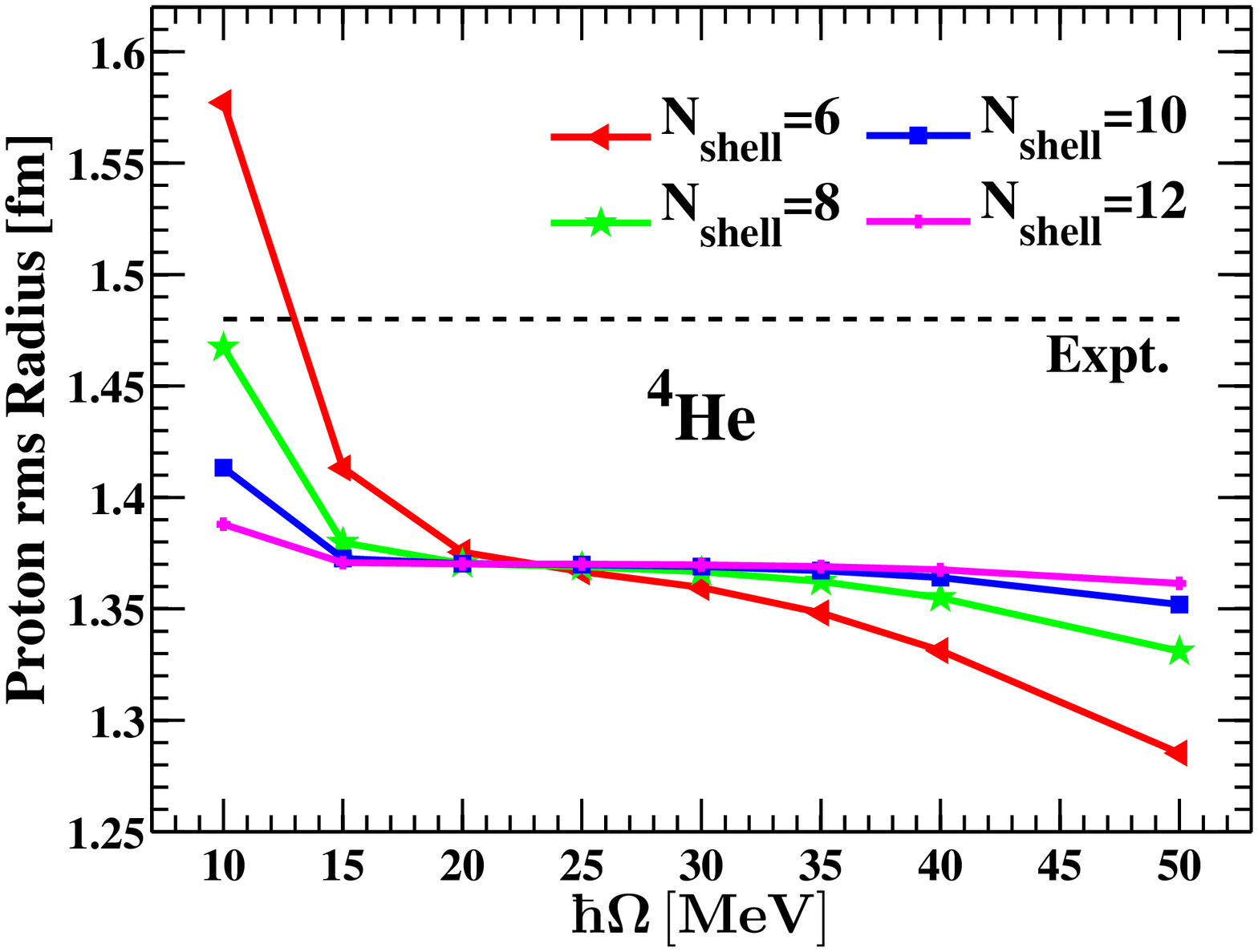}
\caption{\label{fig:he4_jisp_rms} Point-proton rms radius of $^{4}$He as a function of  the oscillator parameter $\hbar \Omega$
for different $N_{\text{shell}}$.
The JISP16 potential \cite{PhysRevC.70.044005,Shirokov200596,Shirokov200733} is used.}
\end{figure}
\begin{figure}
\setlength{\abovecaptionskip}{0pt}
\setlength{\belowcaptionskip}{0pt}
\includegraphics[scale=0.50]{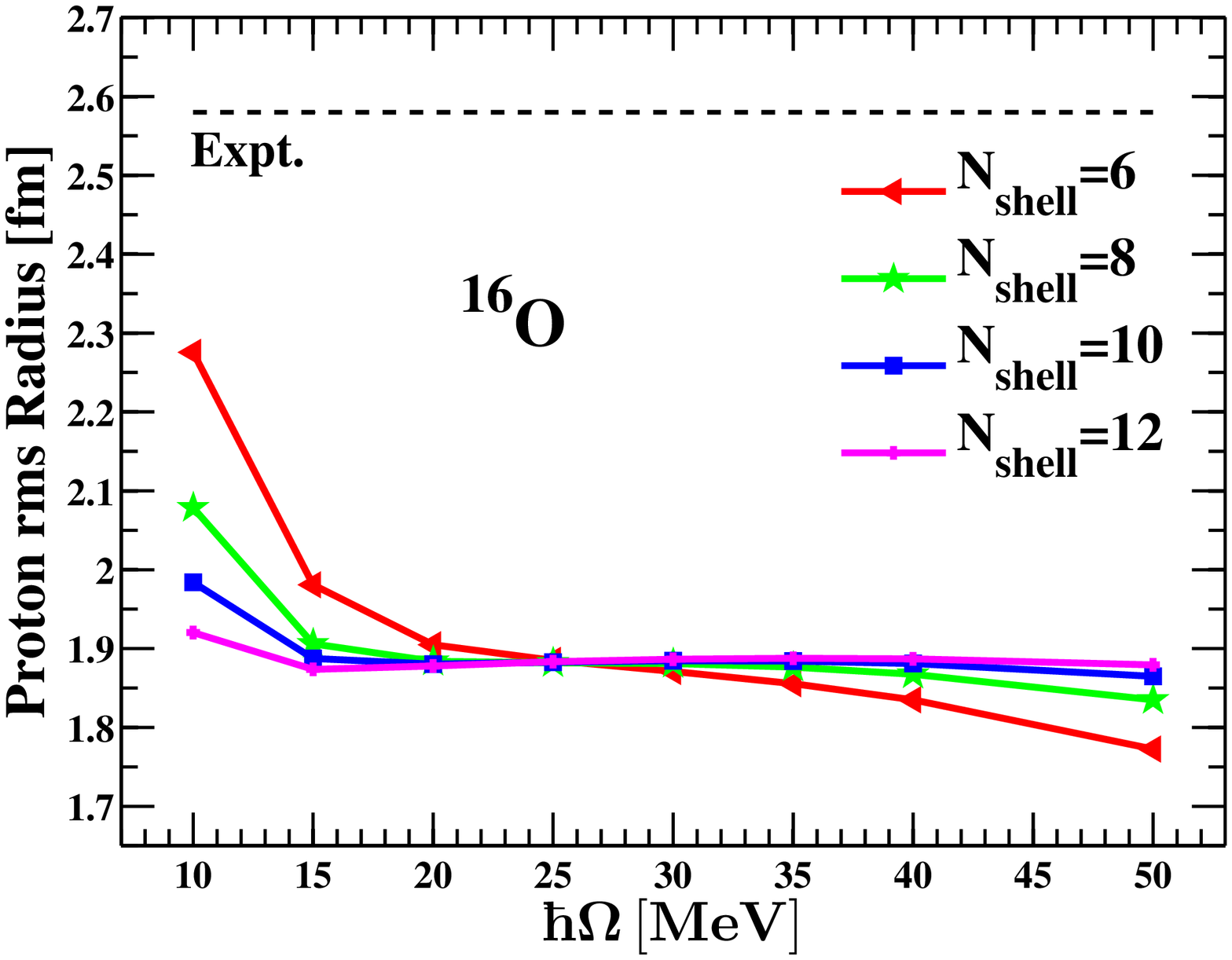}
\caption{\label{fig:o16_jisp_rms}
Point-proton rms radius of $^{16}$O as a function of  the oscillator parameter $\hbar \Omega$
for different $N_{\text{shell}}$.
The JISP16 potential \cite{PhysRevC.70.044005,Shirokov200596,Shirokov200733} is used.}
\end{figure}
\begin{table}
\caption{\label{tab:3}%
Ground-state binding energy and point-proton radius of $^{4}$He with the ``bare'' JISP16 interaction \cite{PhysRevC.70.044005,Shirokov200596,Shirokov200733} at $\hbar \Omega=35$ MeV.
The results of HF-MBPT are obtained with $N_{\text{shell}}=10$.
The NCSM results with $N_{\text{max}}=10$ are taken from Ref.~\cite{PhysRevC.79.014308,Gianina}.
The experimental energy is from Ref.\cite{1674-1137-36-12-001},
 and the experimental radius is obtained as in Table~\ref{tab:n3lo-he4-r}.
}
\begin{ruledtabular}
\begin{tabular}{lcdr}
\textrm{}&
\textrm{Proton rms radius (fm)}&
\multicolumn{1}{c}{\textrm{$E_{\text{\text{g.s.}}}$ (MeV)}}\\
\colrule
 Expt.& 1.477 & -28.296 \\
  NCSM& $1.418$ & -28.222 \\
  SHF           & 1.562 & -22.462 \\
 PT2& 0.015  & -4.373 \\
 PT3& $-$  & -0.803 \\
 $\Delta r_{\text{c.m.}}$ & -0.211 & $-$\\
 HF-MBPT totally& 1.366  & -27.638 \\
\end{tabular}
\end{ruledtabular}
\end{table}
\begin{table}
\caption{\label{tab:4}%
Ground-state binding energy and point-proton radius of $^{16}$O with the ``bare'' JISP16 interaction \cite{PhysRevC.70.044005,Shirokov200596,Shirokov200733} at $\hbar \Omega=35$ MeV.
The results of HF-MBPT are obtained with $N_{\text{shell}}=10$.
The NCSM results with $N_{\text{max}}=8$ are taken from Ref.~\cite{PhysRevC.79.014308,Gianina}.
The experimental energy is from Ref.~\cite{1674-1137-36-12-001},
 and the experimental radius is obtained as in Table~\ref{tab:n3lo-o16-r}.
}
\begin{ruledtabular}
\begin{tabular}{lccr}
\textrm{}&
\textrm{Proton rms radius (fm)}&
\multicolumn{1}{c}{\textrm{$E_{\text{g.s.}}$ (MeV)}}\\
\colrule
 Expt.& 2.581 & -127.619 \\
 NCSM& 1.836&  -131.091 \\
  SHF           & 1.852 & -71.638 \\
 PT2& 0.052  & -58.873 \\
 PT3& $-$  & -4.260 \\
 $\Delta r_{\text{c.m.}}$& -0.061 & $-$ \\
 HF-MBPT totally& 1.843  & -134.771 \\
\end{tabular}
\end{ruledtabular}
\end{table}

Similar to the investigations with the chiral N$^3$LO potential,
we have applied the ``bare'' two-body JISP16 interaction to $^4$He and $^{16}$O.
Figs.~\ref{fig:jispo16}
show calculated binding energies for these two closed-shell nuclei.
Figs.~\ref{fig:he4_jisp_rms} and ~\ref{fig:o16_jisp_rms}
are the radii calculations.
Good convergence is obtained
as indicated by the improved independence of $\hbar \Omega$ and $N_{\text{shell}}$ with increasing $N_{\text{shell}}$.
The JISP16 potential without three-body force gives reasonable ground state energies compared with data.
Tables~\ref{tab:3} and ~\ref{tab:4} give the details of the HF-MBPT calculations with JISP16.
To see how well the HF-MBPT approach does,
we have made a comparison with the benchmark given by the NCSM calculation
with the same JISP16 \cite{PhysRevC.79.014308,Gianina}.
For the NCSM calculation, we introduce the model space truncation parameter $N_{max}$ that
measures the maximal allowed HO excitation energy above the unperturbed lowest zero-order reference state.
We choose to compare out results with $N_{max}$=10 for $^{4}$He calculations,
impling that a total of 11 major HO shells are involved.
Such a model space is sufficient for $^{4}$He.
For the HF-MBPT calculation, fast convergence with increasing the size of
the model space $N_{\text{shell}}$ has been shown in Fig.~\ref{fig:jispo16}.
We use the results of HF-MBPT with $N_{\text{shell}}$=10
to compare with the results of NCSM with $N_{\text{max}}=10$ as in Table~\ref{tab:3}.
We see that HF-MBPT and NCSM calculations give similar results for the energy and radius of $^4$He,
in good agreement with data.
For $^{16}$O, we use $N_{\text{max}}$=8, which corresponds to a total of 10 major HO shells involved.
The results of HF-MBPT with $N_{\text{shell}}$=10 truncation is used to compare with the  NCSM results as in Table~\ref{tab:4}.
Both HF-MBPT and the NCSM give larger binding energies but smaller radii than experimental data.
The MBPT convergence with perturbative order
in the ``bare'' JISP16 calculation
is similar to that in the chiral N$^3$LO calculation.
With the calculations based on N$^3$LO and JISP16,
we may conclude that the MBPT method
can give fairly converged results
in the HF single-particle basis for these realistic $NN$ interactions.

\section{Summary}

We have performed the HF-MBPT calculations with the realistic $NN$ interactions
chiral N$^3$LO and ``bare'' JISP16.
The detailed formulation and anti-symmetrized Goldstone diagram expansions are given.
While the bare N$^3$LO potential is softened using the SRG method,
the ``bare'' JISP16 is employed without softening..
The MBPT corrections are performed based on the spherical Hartree-Fock approach.
The spherical symmetry preserves the quantum numbers of angular momenta.
The angular momentum coupled scheme can significantly
reduce the model dimension and save the computational resources.
As an improvement,
we correct the one-body density for the calculation of the radius
using anti-symmetrized Goldstone diagram expansions through second order.

The closed-shell nuclei, $^4$He and $^{16}$O,
have been chosen as examples for the present HF-MBPT calculations.
Convergence with respect to the SRG-softening parameter,
harmonic oscillator frequency and model space truncation have been discussed in detail.
Our results are consistent
with other works published with MBPT or with other {\it ab initio} methods.
We discussed the MBPT convergence order by order,
showing that corrections up to the third order in energy
and up to the second order in radius
appear to be reasonable when one performs the HF-MBPT calculations
within the Hartree-Fock single-particle basis.
It is demonstrated that smaller contributions from the neglected higher orders decrease with decreasing SRG-softening parameter $\lambda$.
In the present calculations, three-body and higher-order forces are not considered.
To check the convergence of the MBPT calculation,
we have made comparisons with benchmarks given by NCSM calculations with the same $NN$ potential. Consistent results have been obtained.
In general, the calculated radii are smaller than experimental values,
which is a common problem in current {\it ab initio} calculations with these interactions..

\begin{acknowledgments}
Valuable discussions with R. Machleidt and L. Coraggio are gratefully acknowledged.
This work has been supported by
the National Key Basic Research Program of China under Grant No. 2013CB834402;
the National Natural Science Foundation of China under Grants No. 11235001, No. 11320101004 and NO. 11575007;
the CUSTIPEN (China-U.S. Theory Institute for Physics with Exotic Nuclei) funded by the U.S.  Department of Energy, Office of Science under grant number DE-SC0009971;
the Department of Energy under Grant No. DE-FG02-87ER40371;
and National Training Program of Innovation for Undergraduates.
\end{acknowledgments}

\bibliography{references}
\end{document}